\documentclass[aps,prl, twocolumn,superscriptaddress,a4paper,floatfix]{revtex4-2}

\usepackage[utf8]{inputenc}
\usepackage{graphicx}
\usepackage{amsmath,amssymb}
\usepackage[colorlinks=true, linkcolor=black, citecolor=black, urlcolor=blue]{hyperref}
\usepackage{datetime}
\usepackage{braket}
\usepackage{cleveref}
\usepackage{gensymb}
\usepackage{mathtools}
\usepackage{dsfont}
\usepackage{braket}
\usepackage{xcolor}
\usepackage{fdsymbol}
\usepackage{tikz}

\definecolor{devon1}{HTML}{d1cdf6} 
\definecolor{devon2}{HTML}{989be7}
\definecolor{devon3}{HTML}{3569ad}

\definecolor{grey1}{HTML}{d3d3d3}
\definecolor{grey2}{HTML}{808080}

\definecolor{batlow1}{HTML}{1a5862}
\definecolor{batlow2}{HTML}{647e43}
\definecolor{batlow3}{HTML}{efb094}
\definecolor{batlow4}{HTML}{ffd8d6}

\definecolor{orange1}{HTML}{ff8c00} 
\definecolor{devon4}{HTML}{293367} 
\definecolor{devon5}{HTML}{8b94c2} 

\definecolor{devon6}{HTML}{6181d1} 
\definecolor{devon7}{HTML}{bab4f1}

\definecolor{batlowS1}{HTML}{011959}
\definecolor{batlowS2}{HTML}{0e375e}
\definecolor{batlowS3}{HTML}{134c61}

\definecolor{batlowS4}{HTML}{205f61}
\definecolor{batlowS5}{HTML}{396e58}
\definecolor{batlowS6}{HTML}{597a48}

\definecolor{batlowS7}{HTML}{7e8737}
\definecolor{batlowS8}{HTML}{a99635}
\definecolor{batlowS9}{HTML}{d0a35a}

\definecolor{batlowS10}{HTML}{e9ac86}
\definecolor{batlowS11}{HTML}{fac0b5}
\definecolor{batlowS12}{HTML}{ffe4e3}

 \newcommand{\geneva}{Department of Quantum Matter Physics, University of Geneva, Quai Ernest-Ansermet 24, 1211 Geneva, Switzerland}
\newcommand{\EPFL}{Institute of Physics and Center for Quantum Science and Engineering, Ecole Polytechnique F\'ed\'erale de Lausanne (EPFL), CH-1015 Lausanne, Switzerland}
\newcommand{\ETH}{Institute for Theoretical Physics, ETH Zürich, CH-8093 Zürich, Switzerland}
\newcommand{\bonnpi}{Physikalisches Institut, University of Bonn, Nussallee 12, 53115 Bonn, Germany}
\newcommand{\Harvard}{Lyman Laboratory, Department of Physics, Harvard University, Cambridge, MA 02138, USA}

\DeclareRobustCommand\sampleline[1]{%
  \tikz\draw[#1, line width=1.8pt] (0,0) (0,\the\dimexpr\fontdimen22\textfont2\relax)
  -- (1.9em,\the\dimexpr\fontdimen22\textfont2\relax);%
}
\begin{document}

\title{Non-equilibrium dynamics of long-range interacting Fermions}

\author{T.~Zwettler}
\affiliation{\EPFL}
\author{G.~Del Pace}
\affiliation{\EPFL}
\author{F.~Marijanovic}
\affiliation{\ETH}
\author{S.~Chattopadhyay}
\affiliation{\ETH}
\affiliation{\Harvard}
\author{T.~Bühler}
\affiliation{\EPFL}
\author{C.-M.~Halati}
\affiliation{\geneva}
\author{L.~Skolc}
\affiliation{\ETH}
\author{L.~Tolle}
\affiliation{\bonnpi}
\affiliation{\geneva}
\author{V.~Helson}
\affiliation{\EPFL}
\author{G.~Bolognini}
\affiliation{\EPFL}
\author{A.~Fabre}
\affiliation{\EPFL}
\author{S.~Uchino}
\affiliation{Faculty of Science and Engineering, Waseda University, Tokyo 169-8555, Japan}
\author{T.~Giamarchi}
\affiliation{\geneva}
\author{E.~Demler}
\affiliation{\ETH}
\author{J.P.~Brantut}
\affiliation{\EPFL}

\begin{abstract}
    A fundamental problem of out-of-equilibrium physics is the speed at which the order parameter grows upon crossing a phase transition. Here, we investigate the dynamics of ordering in a Fermi gas undergoing a density-wave phase transition induced by quenching of long-range, cavity-mediated interactions. We observe in real-time the exponential rise of the order parameter and track its growth over several orders of magnitude. Remarkably, the growth rate is insensitive to the contact interaction strength from the ideal gas up to the unitary limit and can exceed the Fermi energy by an order of magnitude, in quantitative agreement with a linearized instability analysis. We then generalize our results to linear interaction ramps, where deviations from the adiabatic behaviour are captured by a simple dynamical ansatz. Our study offers a paradigmatic example of the interplay between non-locality and non-equilibrium dynamics, where universal scaling behaviour emerges despite strong interactions at the microscopic level.
\end{abstract}

\maketitle

 Understanding the coherent evolution of many-body systems crossing a phase transition is a ubiquitous problem in physics. Prominent examples are inflation and cosmological phase transitions \cite{baumann2012tasi}, photo-induced states in solid-state physics \cite{Cavalleri:2018aa, delatorre:2021aa}, or adiabatic quantum computing \cite{albash:2018aa}. The central question of study is how fast the order of a symmetry-broken state emerges. 
 In fermionic systems, the Pauli principle plays a crucial role. Already at equilibrium, it makes the non-interacting system prone to instabilities at infinitesimal attractive interactions \cite{ProcEnricoFermi2007, ZwergerTBCS-BECCATUFG2012}, but also very robust against instabilities at large repulsive interactions. For example, while Bose gases can display a rich set of interaction-induced magnetic properties \cite{Stamper-KurnSBGSMAQD2013, BookjansQPTIASBEC2011, SadlerSSBIAQFSBEC2006,strobel:2014ab,luo:2017ab,evrard:2021aa}, the famous Stoner instability that would turn a Fermi gas into a ferromagnet at large short-range repulsion could not be observed in spite of extensive investigations with ultra-cold gases \cite{JoIFIAFGOUA2009, Valtolina:2017aa, Sanner:2011aa, AmicoTROOCAARSRCISIFG2018}. 

The recent advent of fermionic systems with controlled long-range interactions \cite{ De-Marco:2019aa, chen:2023aa, zhang:2021tr, HelsonDWOIAUFGWPMI2023} opens the possibility to study far-from-equilibrium dynamics of order parameters in a non-local setting. In systems with contact interactions the Fermi energy typically sets the fastest evolution timescale. Indeed, in strongly interacting cold-atoms experiments it was observed that fermionic systems relax at a universal rate \cite{Cao:2011ab,Koschorreck:2013aa,Bardon:2014aa,Joseph:2015aa, Luciuk:2017aa,patel:2020th}. Upon crossing a phase transition, systems with local interactions develop order from non-linear defect dynamics after an initial instability seed, while systems with non-local interactions can grow order homogeneously over larger distances and longer times \cite{defenu:2023aa}. Extracting the genuine scaling behaviour governing the non-equilibrium evolution of the unstable system towards order requires following the time-evolution of the order parameter over several decades.

Here, we exploit the unique settings of cavity quantum electrodynamics to record real-time quantum trajectories of a strongly interacting Fermi gas, undergoing density-wave (DW) ordering after a quench of cavity-mediated long-range interactions. In this system, virtual photons in the cavity mediate the interaction, while rare, spontaneous photon leakage events directly carry \textit{in-vivo} information about the order parameter dynamics \cite{Baumann:2010aa,KlinderDPTITODM2015,Ritsch:2013aa,Vaidya:2018aa, mivehvar:2021aa}. We directly observe the exponential growth of order at strong long-range interactions, the hallmark of a dynamical instability, and track it over a wide range of microscopic parameters in the Fermi gas. This reveals a set of universal properties distinct from short-range systems \cite{CetinaUMBIOICTAFS2016, JoIFIAFGOUA2009, PekkerCBPAFIUFGNFR2011}, including a clear, exponential growth of the order parameter over an extended timescale. The growth rate can exceed the Fermi energy by over an order of magnitude, in stark contrast to systems with only contact interactions. Universality is further illustrated by the response to linear ramps of the interaction strength through the transition point. We observe that the growth of the order parameter during the ramp is well captured by a dynamical scaling ansatz which involves only the instantaneous instability rate, as extracted in quench experiments. Thus, the systems response to linear ramps is fully determined, not by its behaviour at criticality, but through its quench properties.

\begin{figure}
\includegraphics{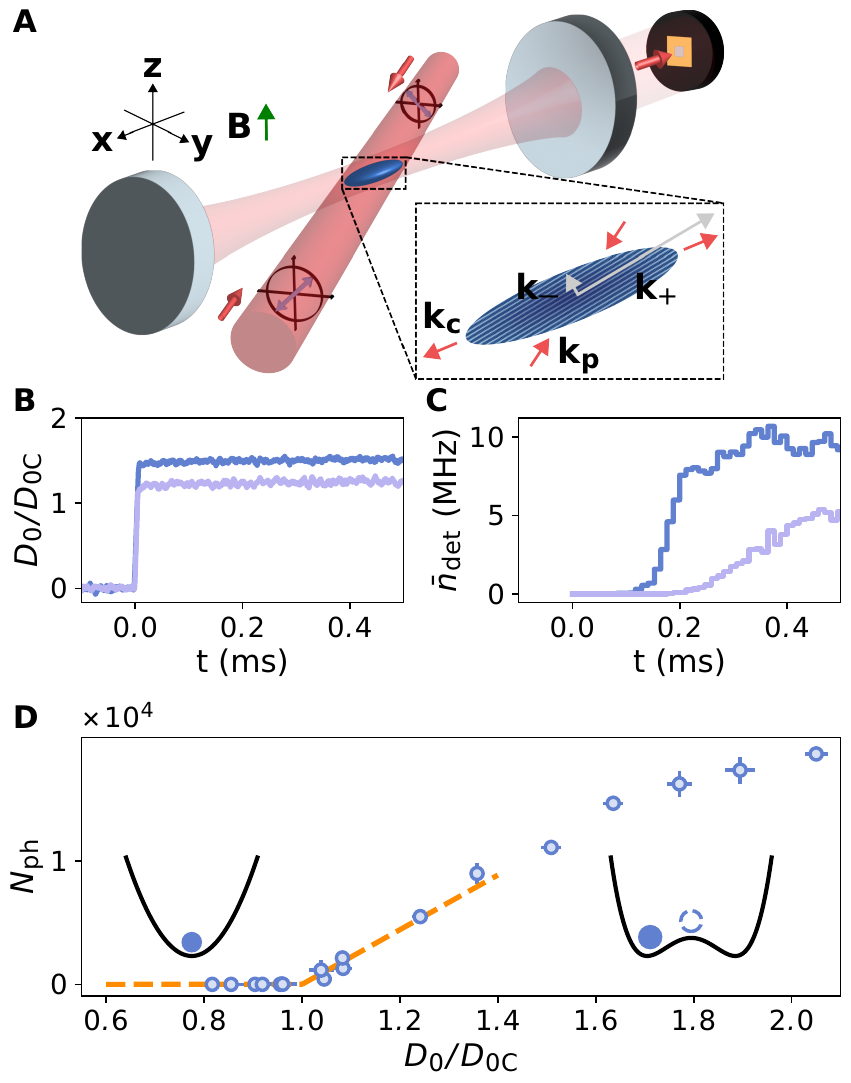}
\caption{\textbf{Concept of the experiment.}
\textbf{(A)} A strongly interacting Fermi gas of ${}^6$Li, residing inside a high-finesse optical resonator, is illuminated by a retro-reflected pump laser with cross-polarization. Pump photons dispersively couple to atomic motion and impart momentum kicks of magnitude $k_{\pm} = |\mathbf{k}_\mathrm{p} \pm \mathbf{k}_\mathrm{c}|$ on atoms through scattering from the pump into the cavity and vice versa.
\textbf{(B)} The long-range interaction strength $D_0$ is quenched across the critical point $D_{0\mathrm{C}}$ by suddenly switching on the power of the pump beam, here shown for $D_0/D_{0\mathrm{C}} = 1.24$ (\textcolor{devon7}{\sampleline{}}) and $1.51$ (\textcolor{devon6}{\sampleline{}}).
\textbf{(C)} Photon flux exiting the cavity measured on a single photon counting module following the quenches displayed in panel B.
\textbf{(D)} Cumulative count of photons $N_{\mathrm{ph}}$ detected during a fixed time window ($ \textcolor{devon6}{\medcircle}$) subsequent to the quench. We identify $D_{0\mathrm{C}}$ by the onset of a finite $N_{\mathrm{ph}}$ and extract its value by a linear fit (\textcolor{orange1}{\sampleline{dashed}}). The critical value coincides with the onset of an instability of the symmetric equilibrium in the free-energy of the system \cite{HelsonDWOIAUFGWPMI2023}.
\label{fig:exp_setup}}
\end{figure}

\subsection{Experimental system}

We produce strong long-range interactions in a unitary Fermi gas by placing the atoms in a high-finesse Fabry-Perot cavity in the presence of side-pumping. Above a critical long-range interaction strength, the system undergoes a quantum phase transition into a DW ordered phase \cite{zhang:2021tr,HelsonDWOIAUFGWPMI2023}, at a wavevector determined by the cavity and pump geometry. When this change in interaction strength occurs suddenly, the system finds itself in an unstable equilibrium and begins to evolve towards an ordered state that minimizes its free energy, as presented in Fig.~\ref{fig:exp_setup}D. The early-time collective dynamics are analogous to an inverted harmonic oscillator, where the imaginary oscillation frequency sets the time scale of the process, a model ubiquitous across fields of physics \cite{SubramanyanPOTIHO2021}. In our case, the single-mode nature of the cavity enforces ordering at all spatial locations simultaneously.

We start the experiments with a spin-balanced, degenerate Fermi gas of ${}^6$Li close to a Feshbach resonance, trapped within the fundamental mode of a high-finesse optical cavity  \cite{Roux:2020aa,roux:2021uf}. The gas has a Fermi energy of $E_{\mathrm{F}} = h \times 20.7\ \mathrm{kHz}$. We illuminate the cloud from the side with a linearly cross-polarized, retro-reflected pump beam as illustrated in Fig.~\ref{fig:exp_setup}A. The pump beam is detuned by $- 2 \pi \times 47.690\ \mathrm{GHz}$ from the atomic $D_2$ transition, but has a close detuning of $\tilde{\Delta}_{\mathrm{c}} \sim \mathrm{MHz}$ from a dispersively-shifted cavity resonance, such that direct Rayleigh scattering in the cavity is off-resonant but second-order photon-exchange processes dominate, yielding an atom-atom interaction extending over the entire mode volume of the cavity. Photon exchanges impart discrete momentum kicks of $k_{\pm} = |\mathbf{k}_{\mathrm{p}} \pm \mathbf{k}_{\mathrm{c}}|$ to the atoms, where $\mathbf{k}_{\mathrm{p,c}}$ are the wave vectors of the pump and cavity photons respectively (see Fig.~\ref{fig:exp_setup}A). The angle of $18^{\circ}$ between pump and cavity axis results in the hierarchy $k_- \ll k_\mathrm{F}\ll k_+$, with the Fermi wave vector $k_\mathrm{F} = \sqrt{2 m E_\mathrm{F}}/\hbar$. The resulting long-range interaction potential between atoms reads (see SI):
\begin{multline}
        \hat{D}(\mathbf{r},\mathbf{r}')=D_0\left[ \cos \left(\mathbf{k}_+ \cdot (\mathbf{r} -\mathbf{r}' ) \right) + \cos \left(\mathbf{k}_- \cdot (\mathbf{r} -\mathbf{r}' ) \right) +\right.\\ \left. \cos \left(\mathbf{k}_+\mathbf{r} + \mathbf{k}_-\mathbf{r}' \right) + \cos \left(\mathbf{k}_-\mathbf{r} + \mathbf{k}_+\mathbf{r}' \right) \right],
        \label{eqn:Interaction}
\end{multline}
with $D_0 \propto V_0/\tilde{\Delta}_{\mathrm{c}}$ and $V_0$ is the light shift induced by the pump beam. In contrast with \cite{zhang:2021tr,wu:2023aa}, we operate in the strong dispersive coupling regime where the photon decay rate $\kappa$ is much smaller than both the dispersive shift and $\tilde{\Delta}_{\mathrm{c}}$, such that dissipation due to cavity-photon losses does not dominate the phase diagram nor the dynamics.

\subsection{Quench dynamics}

We perform a sudden quench of $D_0$ in the atom-cavity system by switching on the pump beam from zero to a finite value at time $t=0$, as presented in Fig.~\ref{fig:exp_setup}B, and record the photon flux $\bar{n}_{\mathrm{det}}$ leaking from the cavity over a single realization using a single photon counter (SPC). The cross-polarization used for the pump beam ensures that there is no optical lattice along the pump direction, such that no atomic dynamics is initiated along the direction $\mathbf{k}_{\mathrm{p}}$. For $D_0 > D_{0\mathrm{C}}$, we observe $\bar{n}_{\mathrm{det}}$ building up in time, as shown for examples in Fig.~\ref{fig:exp_setup}C. In our regime of $E_{\mathrm{F}} \ll \hbar\tilde{\Delta}_{\mathrm{c}}$, there is a direct correspondence between DW order in the gas and the photon field in the cavity, such that photon traces represent individual, real-time quantum trajectories of the ordering process in the cloud. Integrating $\bar{n}_{\mathrm{det}}$ over time following quenches with varying $D_0$ allows to precisely determine the phase boundary $D_{0\mathrm{C}}$, as presented in Fig.~\ref{fig:exp_setup}D. 

\begin{figure}[h]
 \includegraphics{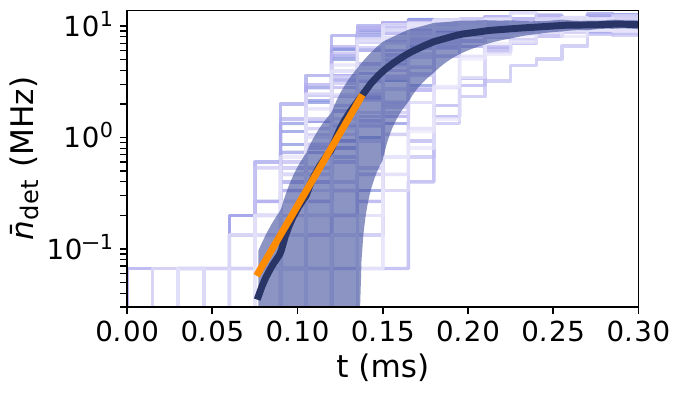}
\caption{\textbf{Exponential growth of the instability.}
Photon flux $\bar{n}_{\mathrm{det}}$ following a sudden long-range interaction quench to $D_0= 1.7\ \times\ D_{0\mathrm{C}}$, averaged over 50 realizations of the experiment. The average photon flux (\textcolor{devon4}{\sampleline{}}) shows an uninterrupted exponential growth over almost two orders of magnitude with the shaded region displaying the standard deviation. The individual photon flux traces are shown as bin-wise step functions, which we fit each with an exponential and deduce a mean growth rate of $60\ \mathrm{kHz}$ (\textcolor{orange1}{\sampleline{}}).
\label{fig:exponential_growth}}
\end{figure}

To understand the dynamics of ordering, we systematically record time traces of $\bar{n}_{\mathrm{det}}$ following a $D_0$ quench. Averaging individual photon traces taken in similar conditions, we reconstruct the time evolution of the cavity-photon number over a large dynamic range. A typical curve is presented in Fig.~\ref{fig:exponential_growth}, for a quench to $1.7\ \times\ D_{0\mathrm{C}}$. We observe an uninterrupted exponential growth over more than one order of magnitude, starting from the earliest time. This is the generic behaviour of ordering at early times in the unstable regime \cite{SadlerSSBIAQFSBEC2006}, which we observe here with a high dynamic range owing to the cavity QED setting. 
We find that the growth of order is well described by an exponential for all the parameters explored. It is however remarkable that the exponential growth persists over more than a decade, without any indication of a change of behaviour until reaching the largest values. 
The long-range character of the interactions prevents the proliferation of defects in the ordering process, such that the runaway evolution of the unstable oscillator can proceed without yielding any defect coarsening dynamics. As a result, the ordering is described by a single growth rate $\alpha$. We find that this behaviour in not restricted to our bulk, three-dimensional Fermi gas but is also observed in numerical simulations for one-dimensional systems coupled to a classical cavity mode (see SI). 

\begin{figure*}[t!]
 \includegraphics{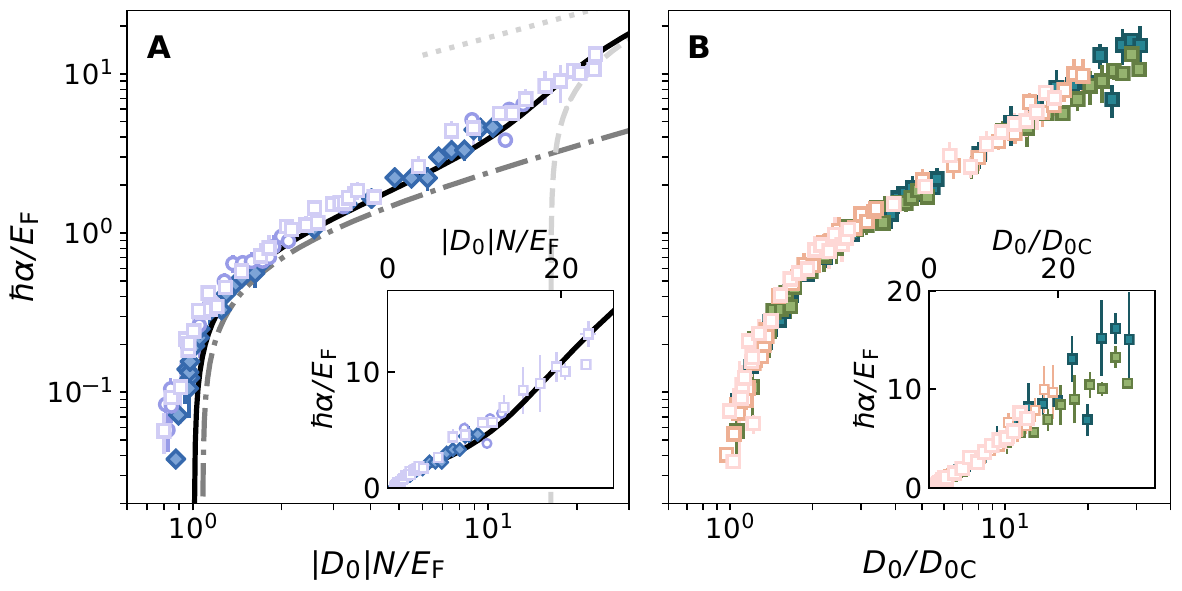}
\caption{\textbf{Instability growth rates.}
\textbf{(A)} Instability growth rate $\alpha$ in units of the Fermi energy $E_{\mathrm{F}}$ at unitarity as a function of $D_0$ for different pump-cavity detunings $\tilde{\Delta}_{\mathrm{c}}$: $-1.4\ \mathrm{MHz}$ ($ \textcolor{devon1}{\medsquare}$), $-2.5\ \mathrm{MHz}$ ($ \textcolor{devon2}{\medcircle}$) and $-3.4\ \mathrm{MHz}$ ($ \textcolor{devon3}{\meddiamond}$). Theoretical prediction based on an instability analysis including trap-averaging (\sampleline{}), with the individual contributions of $k_{+}$ (\textcolor{grey1}{\sampleline{dashed}}) and $k_{-}$ (\textcolor{grey2}{\sampleline{dash pattern=on .7em off .2em on .05em off .2em}}). $\alpha$ converges to a limiting value given by the f-sum rule (\textcolor{grey1}{\sampleline{dotted}}) for $D_0\longrightarrow \infty$.
\textbf{(B)} Growth rates in the BEC-BCS crossover at $1/k_{\mathrm{F}}a$: $1.1$ ($ \textcolor{batlow1}{\medsquare}$), $0$ ($ \textcolor{batlow2}{\medsquare}$), $-0.7$ ($ \textcolor{batlow3}{\medsquare}$) for $\tilde{\Delta}_{\mathrm{c}}=-1.4\ \mathrm{MHz}$, and for a non-interacting, polarized Fermi gas ($ \textcolor{batlow4}{\medsquare}$), for $-0.8\ \mathrm{MHz}$ (see SI for extended data). For each curve, $D_0$ is normalized by the value $D_{0\mathrm{C}}$ observed for the corresponding interaction strength. The inset of both panels report the same data as the main figure in linear scale.
\label{fig:growth_rate}}
\end{figure*}

We now proceed to systematically explore the evolution of $\alpha$ in a unitary Fermi gas quenched into the DW ordered phase, for $D_{0}$ up to $30\ \times\  D_{0\mathrm{C}}$. The growth rate is obtained by a fit of
individual photon flux traces. The results are presented in Fig.~\ref{fig:growth_rate}A for three different choices of $\tilde{\Delta}_{\mathrm{c}}$. The collapse observed for these three detunings demonstrates that the only relevant parameter is the long-range interaction strength. For $D_0 $ approaching $D_{0\mathrm{C}}$, $\alpha$ tends to zero, which is a manifestation of the critical slowing down close to the phase transition \cite{defenu:2023aa}. Deeper in the ordered phase, $\alpha$ monotonically increases with $D_0$ over two orders of magnitude, showing no sign of saturation. It reaches and eventually exceeds the Fermi energy by about an order of magnitude for the largest $D_0$, meaning that the system orders much faster than the local fermionic timescale. Effectively, the long-range character of the interactions circumvents the effects of the Pauli exclusion on interactions.

To uncover the microscopic origin of this behaviour, we model the early-time evolution following the quench using a linear-response theory of the instability (see SI). This framework amounts to a linear stability analysis, naturally yielding an exponential growth of the order parameter for the unstable polaritonic mode. This growth is seeded by initial fermionic density fluctuations in the system and coherently amplified by the unitary dynamics. This approach leads to the implicit equation for $\alpha$:
\begin{equation}
   1 = \frac{D_0 N}{8} \sum_{q= \pm k_{\pm}} \int_{-\infty}^{\infty}\frac{d\omega}{\pi}\frac{\omega\mathrm{Im}[\chi^R(q, \omega)]}{\alpha^2+\omega^2},
   \label{eq:theory}
\end{equation}
with $\chi^{\mathrm{R}}(q, \omega)$ the retarded density-density response function at momentum $q$ and frequency $\omega$, evaluated prior to the quench for the equilibrium gas at unitarity and including averaging over the harmonic trap. The natural scale of $\chi^{\mathrm{R}}$ is the Fermi energy, motivating the normalization of the data by $E_\mathrm{F}$ in Fig.~\ref{fig:growth_rate}. Using the random-phase-approximation result for $\chi^{\mathrm{R}}(k_\pm, \omega)$ at unitarity \cite{he:2016aa}, we obtain a prediction for the growth rate without free parameters for all values of $D_0$, plotted in Fig. \ref{fig:growth_rate}A as solid black line. The agreement is very good over all the range covered. This analysis also highlights the role of the modes at $k_-$ and $k_+$ in the dynamics: the low-energy $k_-$ mode controls the critical point and the dynamics close to it, while the high-energy mode at $k_+$ dominates the growth of the order parameter for very large $D_0$, away from criticality. The specific contributions of modes $k_\pm$ are presented in dashed and dotted-dashed lines in Fig.~\ref{fig:growth_rate}A. Because $k_+ \gg k_-$, we observe a crossover between the $k_-$-dominated dynamics close to the critical point and the $k_+$-dominated dynamics at very large $D_0$.

The absence of a growth rate saturation at the Fermi energy and the consistent occurrence of the exponential growth over the whole parameter range suggests that the microscopic properties of the gas do not play a crucial role. To further confirm this intuition, we reproduce the experiments away from unitarity through the crossover between Bose-Einstein condensate (BEC) and Bardeen-Cooper-Schrieffer (BCS) superfluids for finite positive and negative scattering length $a$, as well as for an ideal, fully polarized Fermi gas. The results are presented in Fig.~\ref{fig:growth_rate}B. Upon normalization by $D_{0\mathrm{C}}$, the measured growth rates in the three regimes agree within our signal-to-noise ratio with the measurements performed at unitarity, over the two decades in growth rate explored by the system. This even generalizes to a fully-polarized Fermi gas, without any contact interaction. This striking similarity demonstrates the universal character of the growth rate scaling with long-range interaction strength.

Our theoretical model suggests a possible set of explanation for this remarkable feature. For $\alpha\to \infty$, Eq.~(\ref{eq:theory}) simplifies to the $f$-sum rule, which enforces a universal, interaction and quantum-statistics independent growth rate in the asymptotic regime. The corresponding value is presented in Fig.~\ref{fig:growth_rate} as a dashed line, and is not reached in our experiments. For lower rates, deviations from the sum rule results appear due to the weak dependence of $\chi^R(k_+,\omega)$ on interactions. For $D_0$ closest to $ D_{0\mathrm{C}}$, the growth rate is controlled by the critical behaviour of the system, primarily determined by the zero-frequency response function at $k_-$, which approaches the compressibility for $k_- \ll k_\mathrm{F}$ \cite{helson:2023aa}. In our case, the averaging over the harmonic trap smoothens out the moderate quantitative differences to a range comparable with the experimental noise (see SI for details).

\subsection{Ramp dynamics}

\begin{figure*}[ht!]
 \includegraphics{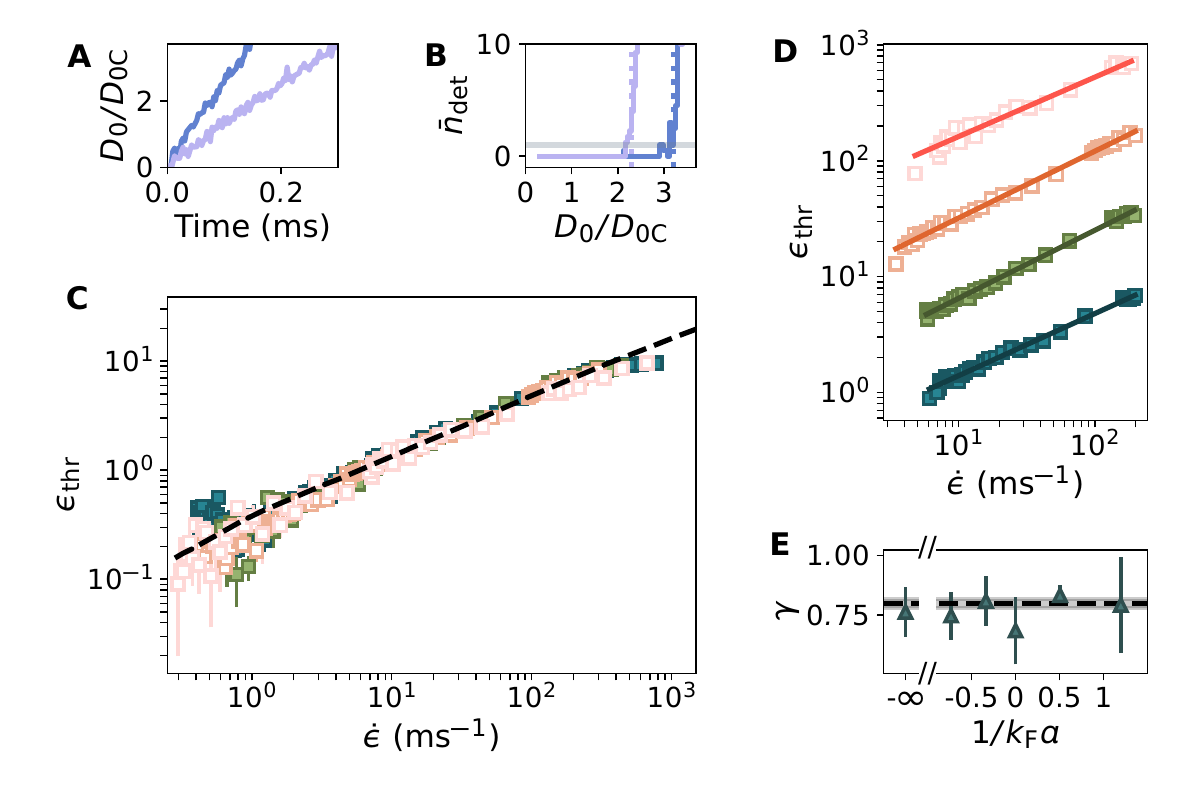}
\caption{\textbf{Linear ramps of long-range interactions.}
\textbf{(A)} Linear ramps of $D_0$ with different speed  of $\dot{\epsilon} = 10.7\,  \mathrm{ms}^{-1}$ (\textcolor{devon7}{\sampleline{}}) and $21.7\, \mathrm{ms}^{-1}$ (\textcolor{devon6}{\sampleline{}}). Each line represents the pump power ramp as experimentally measured.
\textbf{(B)} Cavity photon leakage $\bar{n}_{\mathrm{det}}$ during the ramp, from which we extract the long-range interaction strength $D_{0,\mathrm{thr}}$ (\textcolor{devon7}{\sampleline{dotted}}, \textcolor{devon6}{\sampleline{dotted}}) to reach a threshold count rate of $1\ \mathrm{MHz}$ (\textcolor{grey1}{\sampleline{}}).
\textbf{(C)} Relative difference $\epsilon_{\mathrm{thr}}$ of threshold to the critical point as a function of the quench rate $\dot{\epsilon}$ for $\tilde{\Delta}_{\mathrm{c}}=-1.4\ \mathrm{MHz}$ at $1/k_{\mathrm{F}}a$: $1.1$ ($ \textcolor{batlow1}{\medsquare}$), $0$ ($ \textcolor{batlow2}{\medsquare}$), $-0.7$ ($ \textcolor{batlow3}{\medsquare}$) and for the polarized Fermi gas ($ \textcolor{batlow4}{\medsquare}$). (\sampleline{dashed}) shows the behaviour as a function of the ramp speed predicted from the dynamical ansatz, based on the data of Fig.~\ref{fig:growth_rate} (see text for details). 
\textbf{(D)} Power-law fits to $\epsilon_{\mathrm{thr}}$ for the different interaction strengths reported in panel C, offset by a multiplicative factor of $5$ for clarity.
\textbf{(E)} Power-law exponents for different short-range interaction strengths $1/k_{\mathrm{F}}a$ ($\textcolor{grey2}{\medtriangleup}$), where the point at $-\infty$ represents the polarized, non-interacting Fermi gas. Each data point corresponds to the weighted average, and the error bar the standard deviation among the exponents $\gamma$ obtained over datasets at different $\tilde{\Delta}_{\mathrm{c}}$. We also indicate the critical exponent determined from fitting the dynamical ansatz (\sampleline{dashed}), together with its fit error reported as shaded region.
\label{fig:ramps}}
\end{figure*}

The response of the system to a sudden quench shows that the dynamics over the entire accessible parameter range follow a simple exponential scaling law controlled by $\alpha$. We now extend the study beyond instantaneous quenches to linear ramps of $D_0$ performed at finite speed $\dot{\epsilon}$ with $\epsilon = \frac{D_0 - D_{0\mathrm{C}}}{D_{0\mathrm{C}}}$, as shown in Fig.~\ref{fig:ramps}A. By varying the speed, this protocol interpolates between the adiabatic regime extensively studied in the literature \cite{mivehvar:2021aa} and the sudden quenches observed above. Throughout the linear ramp, time directly maps onto $D_0$, which allows to identify for each $\dot{\epsilon}$ an apparent ordering threshold $D_{0,\mathrm{thr}}$, as presented in Fig.~\ref{fig:ramps}B (see SI for details). We use the photon traces to extract the relative differences of the apparent threshold to the critical value $\epsilon_\mathrm{thr} = \frac{D_{0,\mathrm{thr}} - D_{0\mathrm{C}}}{D_{0\mathrm{C}}}$ over four decades in speed. The results are presented in Fig.~\ref{fig:ramps}C. As the speed increases, we observe an increasing difference in the apparent ordering threshold. This is because at higher speeds, the system orders at later relative times during the ramp. The smooth variations spanning up to a factor of ten in relative increase of the threshold are identical from non-interacting to strongly interacting Fermions across the BEC to BCS crossover. For the lowest speeds, we observe a larger noise in $\epsilon_\mathrm{thr}$ and systematic deviations due to losses of atoms during the ramp. 

Over more than one order of magnitude in speed, $\epsilon_\mathrm{thr}$ exhibits a consistent power-law behaviour. We substantiate this by fitting a power law to the different curves at unitarity, finite $a$ and for an ideal gas, as presented in Fig.~\ref{fig:ramps}D. Such a power-law behaviour could suggest an explanation in terms of a generalized Kibble-Zurek scenario \cite{Baumann:2011vy,KlinderDPTITODM2015,defenu:2023aa}, in which the instantaneous growth rate of the order parameter is integrated throughout the ramp protocol \cite{lamacraft:2007aa}. We fit the power-law behaviour with $\epsilon_{\mathrm{thr}} \propto \dot{\epsilon}^{1/(1+\gamma)}$, as presented in Fig.~\ref{fig:ramps}D. In the Kibble-Zurek framework, $\gamma$ corresponds to the dynamical critical exponent characterizing the phase transition. We find an exponent of 0.731(16), strongly deviating from the value of $0.18$ measured in \cite{KlinderDPTITODM2015} for the open Dicke model with rising pump ramps, but in agreement with the value of $0.75$ found there for decreasing ramps. It also differs from predicted mean-field values for closed ($\gamma=0.5$) and open quantum systems ($\gamma=1$) \cite{Bhaseen:2012aa,acevedo:2014aa,defenu:2018aa,defenu:2023aa}.
 
Rather than connecting the ramp response to the critical behaviour as in the Kibble-Zurek scenario, we can explain the observed scaling law based on the knowledge of the growth rate after a quench. Knowing the exponential scaling law of Fig.~\ref{fig:growth_rate} for a given $D_0$ offers the possibility to compare the ramp speed with the instantaneous growth rate at each point in time, thus quantifying the deviations from the adiabatic behaviour. This leads us to a first-order differential equation ansatz for the photon flux leaking from the cavity as a function of time in presence of time-varying $D_0$: 
\begin{equation}
    \frac{\mathrm{d}\bar{n}_{\mathrm{det}}}{\mathrm{d}t} = \alpha(t)\bar{n}_{\mathrm{det}}(t),
    \label{eq:eqdiff}
\end{equation}
where we propose the scaling ansatz $\alpha(t)=\alpha\left[ D_0(t)\right]$. It generalizes the exponential law while keeping the dynamics independent of any microscopic detail beyond the determination of $\alpha$. Direct integration of Eq.~\eqref{eq:eqdiff} using the data of Fig.~\ref{fig:growth_rate}B yields a prediction for the photon flux throughout the ramps, presented as dashed line in Fig.~\ref{fig:ramps}C. There, we have combined all the data of Fig.~\ref{fig:growth_rate}B together, since they do not show any difference within our signal-to-noise ratio. The agreement is excellent over most of the range, except for the largest speeds, further demonstrating the universal character of the dynamical properties. Interestingly, our scaling ansatz does not tie the scaling to the critical regime as in the Kibble-Zurek picture, but rather to the growth rate of the order parameter. A manifestation of this feature is the role of the high-energy mode at $k_+$, as observed in Fig.~\ref{fig:growth_rate}, which is negligible for the determination of the critical point. 

\subsection{Conclusion and perspectives}

In conclusion, our thorough investigation of the ordering dynamics of long-range interacting Fermions suggests a distinct classification for these systems compared to their short-range interacting counterparts. Here, the entire ordering process follows an exponential growth mechanism, weakly dependent on microscopic details. In short-ranged systems, the ramp protocol reflects the critical properties of the transition, as given by Kibble-Zurek prescriptions. However, in our setting, the long-ranged photon-mediated interactions enforces an infinite correlation length for the order parameter throughout the transition, nullifying the association between diverging spatial and temporal scales at criticality. 

Beyond our theoretical analysis of the three dimensional system, these general conclusions are also supported by our numerical simulations on simplified models in one and two-dimension (see SI). Interestingly, we did not observe photon number ringing in the early time dynamics, nor a late-time steady state regime with high temperature throughout the parameter space explored in our experiments, in contrast with \cite{wu:2023aa}. There, the sudden quench protocol was investigated in a regime of weak collective coupling and $\tilde\Delta_c \sim \kappa$, which indicates that our general framework is not sufficient to explain late-time dynamics in the regime of strong dissipation \cite{schutz:2014aa,schutz:2016aa}. Beyond the context of cavity-QED, our results also provide a framework for the description of Fermi gases interacting at finite range due to dipolar interactions or Rydberg dressing, where the interplay between the finite interaction range and the system size would give rise to a richer scenario. 

By virtue of our cavity-QED system's ability to record not only average time-evolution but single quantum trajectories, we have access to finer information in the noise and correlations of such signals. There, we anticipate that quantum effects, such as measurement back-action from the cavity and quantum noise in the gas, could arise in the form of deviations from the mean-field description.

\section*{ACKNOWLEDGEMENTS}

We thank Simon J\"ager for discussions. The EPFL group acknowledges funding from the Swiss State Secretariat for Education, Research and Innovation (Grants No. MB22.00063 and UeM019-5.1). A.F. acknowledges funding from the EPFL Center for Quantum Science and Engineering. The ETH group acknowledges funding from the the Swiss National Science Foundation project 200021\_212899, the Swiss State Secretariat for Education, Research and Innovation (contract number UeM019-1), and NCCR SPIN. E.D. acknowledges funding from ARO grant number W911NF-21-1-0184. S.C. is grateful for support from the NSF under Grant No. DGE-1845298 \& for the hospitality of the Institute of Theoretical Physics at ETH, Zürich.
S.U. acknowledges funding from by JST PRESTO (JPMJPR235)
and JSPS KAKENHI (JP21K03436).
T.G. and C.-M.H.~acknowledge support by the Swiss National Science Foundation under Division II grants 200020-188687 and 200020-219400, and would like to thank the Institut Henri Poincaré (UAR 839 CNRS-Sorbonne Université) and the LabEx CARMIN (ANR-10-LABX-59-01) for their support.
C.-M.H.~acknowledges support in part by grant NSF PHY-1748958 to the Kavli Institute for Theoretical Physics (KITP).
L.T.~acknowledges support by the Deutsche Forschungsgemeinschaft  (DFG, German  Research  Foundation) under project number 277625399 - TRR 185 (B4), project number 277146847 - CRC 1238 (C05) and under Germany’s Excellence Strategy – Cluster of Excellence Matter and Light for Quantum Computing (ML4Q) EXC 2004/1 – 390534769.

\section*{AUTHORS CONTRIBUTION}

TZ, GDP, TB, VH, GB and AF performed experimental work, FM, SC, CMH, LS, LT, SU performed analytic and numerical calculations. TG, ED and JPB supervised the theoretical and experimental work, respectively. All the authors contributed to the discussion and interpretation of the results. 

\bibliography{soquenchbibliography}

\begin{thebibliography}{64}%
\makeatletter
\providecommand \@ifxundefined [1]{%
 \@ifx{#1\undefined}
}%
\providecommand \@ifnum [1]{%
 \ifnum #1\expandafter \@firstoftwo
 \else \expandafter \@secondoftwo
 \fi
}%
\providecommand \@ifx [1]{%
 \ifx #1\expandafter \@firstoftwo
 \else \expandafter \@secondoftwo
 \fi
}%
\providecommand \natexlab [1]{#1}%
\providecommand \enquote  [1]{``#1''}%
\providecommand \bibnamefont  [1]{#1}%
\providecommand \bibfnamefont [1]{#1}%
\providecommand \citenamefont [1]{#1}%
\providecommand \href@noop [0]{\@secondoftwo}%
\providecommand \href [0]{\begingroup \@sanitize@url \@href}%
\providecommand \@href[1]{\@@startlink{#1}\@@href}%
\providecommand \@@href[1]{\endgroup#1\@@endlink}%
\providecommand \@sanitize@url [0]{\catcode `\\12\catcode `\$12\catcode
  `\&12\catcode `\#12\catcode `\^12\catcode `\_12\catcode `\%12\relax}%
\providecommand \@@startlink[1]{}%
\providecommand \@@endlink[0]{}%
\providecommand \url  [0]{\begingroup\@sanitize@url \@url }%
\providecommand \@url [1]{\endgroup\@href {#1}{\urlprefix }}%
\providecommand \urlprefix  [0]{URL }%
\providecommand \Eprint [0]{\href }%
\providecommand \doibase [0]{https://doi.org/}%
\providecommand \selectlanguage [0]{\@gobble}%
\providecommand \bibinfo  [0]{\@secondoftwo}%
\providecommand \bibfield  [0]{\@secondoftwo}%
\providecommand \translation [1]{[#1]}%
\providecommand \BibitemOpen [0]{}%
\providecommand \bibitemStop [0]{}%
\providecommand \bibitemNoStop [0]{.\EOS\space}%
\providecommand \EOS [0]{\spacefactor3000\relax}%
\providecommand \BibitemShut  [1]{\csname bibitem#1\endcsname}%
\let\auto@bib@innerbib\@empty
\bibitem [{\citenamefont {Baumann}(2012)}]{baumann2012tasi}%
  \BibitemOpen
  \bibfield  {author} {\bibinfo {author} {\bibfnamefont {D.}~\bibnamefont
  {Baumann}},\ }\href@noop {} {\bibinfo {title} {Tasi lectures on inflation}}
  (\bibinfo {year} {2012}),\ \Eprint {https://arxiv.org/abs/0907.5424}
  {arXiv:0907.5424 [hep-th]} \BibitemShut {NoStop}%
\bibitem [{\citenamefont {Cavalleri}(2018)}]{Cavalleri:2018aa}%
  \BibitemOpen
  \bibfield  {author} {\bibinfo {author} {\bibfnamefont {A.}~\bibnamefont
  {Cavalleri}},\ }\bibfield  {title} {\bibinfo {title} {Photo-induced
  superconductivity},\ }\href {https://doi.org/10.1080/00107514.2017.1406623}
  {\bibfield  {journal} {\bibinfo  {journal} {Contemporary Physics}\ }\textbf
  {\bibinfo {volume} {59}},\ \bibinfo {pages} {31} (\bibinfo {year}
  {2018})}\BibitemShut {NoStop}%
\bibitem [{\citenamefont {de~la Torre}\ \emph {et~al.}(2021)\citenamefont
  {de~la Torre}, \citenamefont {Kennes}, \citenamefont {Claassen},
  \citenamefont {Gerber}, \citenamefont {McIver},\ and\ \citenamefont
  {Sentef}}]{delatorre:2021aa}%
  \BibitemOpen
  \bibfield  {author} {\bibinfo {author} {\bibfnamefont {A.}~\bibnamefont
  {de~la Torre}}, \bibinfo {author} {\bibfnamefont {D.~M.}\ \bibnamefont
  {Kennes}}, \bibinfo {author} {\bibfnamefont {M.}~\bibnamefont {Claassen}},
  \bibinfo {author} {\bibfnamefont {S.}~\bibnamefont {Gerber}}, \bibinfo
  {author} {\bibfnamefont {J.~W.}\ \bibnamefont {McIver}},\ and\ \bibinfo
  {author} {\bibfnamefont {M.~A.}\ \bibnamefont {Sentef}},\ }\bibfield  {title}
  {\bibinfo {title} {Colloquium: Nonthermal pathways to ultrafast control in
  quantum materials},\ }\href {https://doi.org/10.1103/RevModPhys.93.041002}
  {\bibfield  {journal} {\bibinfo  {journal} {Rev. Mod. Phys.}\ }\textbf
  {\bibinfo {volume} {93}},\ \bibinfo {pages} {041002} (\bibinfo {year}
  {2021})}\BibitemShut {NoStop}%
\bibitem [{\citenamefont {Albash}\ and\ \citenamefont
  {Lidar}(2018)}]{albash:2018aa}%
  \BibitemOpen
  \bibfield  {author} {\bibinfo {author} {\bibfnamefont {T.}~\bibnamefont
  {Albash}}\ and\ \bibinfo {author} {\bibfnamefont {D.~A.}\ \bibnamefont
  {Lidar}},\ }\bibfield  {title} {\bibinfo {title} {Adiabatic quantum
  computation},\ }\href {https://doi.org/10.1103/RevModPhys.90.015002}
  {\bibfield  {journal} {\bibinfo  {journal} {Rev. Mod. Phys.}\ }\textbf
  {\bibinfo {volume} {90}},\ \bibinfo {pages} {015002} (\bibinfo {year}
  {2018})}\BibitemShut {NoStop}%
\bibitem [{\citenamefont {M.~Inguscio}(2007)}]{ProcEnricoFermi2007}%
  \BibitemOpen
  \bibinfo {editor} {\bibfnamefont {C.~S.}\ \bibnamefont {M.~Inguscio},
  \bibfnamefont {W.~Ketterle}},\ ed.,\ \href@noop {} {\emph {\bibinfo {title}
  {Proceedings, International School of Physics Enrico Fermi, Course CLXIV,
  "Ultra-cold Fermi Gases"}}}\ (\bibinfo  {publisher} {IOS, Amsterdam},\
  \bibinfo {year} {2007})\BibitemShut {NoStop}%
\bibitem [{\citenamefont {Zwerger}(2012)}]{ZwergerTBCS-BECCATUFG2012}%
  \BibitemOpen
  \bibfield  {author} {\bibinfo {author} {\bibfnamefont {W.}~\bibnamefont
  {Zwerger}},\ }\href@noop {} {\emph {\bibinfo {title} {The BCS-BEC crossover
  and the unitary Fermi gas}}},\ \bibinfo {series} {Lecture notes in physics},
  Vol.\ \bibinfo {volume} {836}\ (\bibinfo  {publisher} {Springer},\ \bibinfo
  {address} {Berlin},\ \bibinfo {year} {2012})\BibitemShut {NoStop}%
\bibitem [{\citenamefont {Stamper-Kurn}\ and\ \citenamefont
  {Ueda}(2013)}]{Stamper-KurnSBGSMAQD2013}%
  \BibitemOpen
  \bibfield  {author} {\bibinfo {author} {\bibfnamefont {D.~M.}\ \bibnamefont
  {Stamper-Kurn}}\ and\ \bibinfo {author} {\bibfnamefont {M.}~\bibnamefont
  {Ueda}},\ }\bibfield  {title} {\bibinfo {title} {Spinor bose gases:
  Symmetries, magnetism, and quantum dynamics},\ }\href
  {https://doi.org/10.1103/RevModPhys.85.1191} {\bibfield  {journal} {\bibinfo
  {journal} {Rev. Mod. Phys.}\ }\textbf {\bibinfo {volume} {85}},\ \bibinfo
  {pages} {1191} (\bibinfo {year} {2013})}\BibitemShut {NoStop}%
\bibitem [{\citenamefont {Bookjans}\ \emph {et~al.}(2011)\citenamefont
  {Bookjans}, \citenamefont {Vinit},\ and\ \citenamefont
  {Raman}}]{BookjansQPTIASBEC2011}%
  \BibitemOpen
  \bibfield  {author} {\bibinfo {author} {\bibfnamefont {E.~M.}\ \bibnamefont
  {Bookjans}}, \bibinfo {author} {\bibfnamefont {A.}~\bibnamefont {Vinit}},\
  and\ \bibinfo {author} {\bibfnamefont {C.}~\bibnamefont {Raman}},\ }\bibfield
   {title} {\bibinfo {title} {Quantum phase transition in an antiferromagnetic
  spinor bose-einstein condensate},\ }\href
  {https://doi.org/10.1103/PhysRevLett.107.195306} {\bibfield  {journal}
  {\bibinfo  {journal} {Phys. Rev. Lett.}\ }\textbf {\bibinfo {volume} {107}},\
  \bibinfo {pages} {195306} (\bibinfo {year} {2011})}\BibitemShut {NoStop}%
\bibitem [{\citenamefont {Sadler}\ \emph {et~al.}(2006)\citenamefont {Sadler},
  \citenamefont {Higbie}, \citenamefont {Leslie}, \citenamefont
  {Vengalattore},\ and\ \citenamefont {Stamper-Kurn}}]{SadlerSSBIAQFSBEC2006}%
  \BibitemOpen
  \bibfield  {author} {\bibinfo {author} {\bibfnamefont {L.~E.}\ \bibnamefont
  {Sadler}}, \bibinfo {author} {\bibfnamefont {J.~M.}\ \bibnamefont {Higbie}},
  \bibinfo {author} {\bibfnamefont {S.~R.}\ \bibnamefont {Leslie}}, \bibinfo
  {author} {\bibfnamefont {M.}~\bibnamefont {Vengalattore}},\ and\ \bibinfo
  {author} {\bibfnamefont {D.~M.}\ \bibnamefont {Stamper-Kurn}},\ }\bibfield
  {title} {\bibinfo {title} {Spontaneous symmetry breaking in a quenched
  ferromagnetic spinor bose--einstein condensate},\ }\href
  {https://doi.org/10.1038/nature05094} {\bibfield  {journal} {\bibinfo
  {journal} {Nature}\ }\textbf {\bibinfo {volume} {443}},\ \bibinfo {pages}
  {312} (\bibinfo {year} {2006})}\BibitemShut {NoStop}%
\bibitem [{\citenamefont {Strobel}\ \emph {et~al.}(2014)\citenamefont
  {Strobel}, \citenamefont {Muessel}, \citenamefont {Linnemann}, \citenamefont
  {Zibold}, \citenamefont {Hume}, \citenamefont {Pezz{\`e}}, \citenamefont
  {Smerzi},\ and\ \citenamefont {Oberthaler}}]{strobel:2014ab}%
  \BibitemOpen
  \bibfield  {author} {\bibinfo {author} {\bibfnamefont {H.}~\bibnamefont
  {Strobel}}, \bibinfo {author} {\bibfnamefont {W.}~\bibnamefont {Muessel}},
  \bibinfo {author} {\bibfnamefont {D.}~\bibnamefont {Linnemann}}, \bibinfo
  {author} {\bibfnamefont {T.}~\bibnamefont {Zibold}}, \bibinfo {author}
  {\bibfnamefont {D.~B.}\ \bibnamefont {Hume}}, \bibinfo {author}
  {\bibfnamefont {L.}~\bibnamefont {Pezz{\`e}}}, \bibinfo {author}
  {\bibfnamefont {A.}~\bibnamefont {Smerzi}},\ and\ \bibinfo {author}
  {\bibfnamefont {M.~K.}\ \bibnamefont {Oberthaler}},\ }\bibfield  {title}
  {\bibinfo {title} {Fisher information and entanglement of non-gaussian spin
  states},\ }\href {https://www.science.org/doi/abs/10.1126/science.1250147}
  {\bibfield  {journal} {\bibinfo  {journal} {Science}\ }\textbf {\bibinfo
  {volume} {345}},\ \bibinfo {pages} {424} (\bibinfo {year}
  {2014})}\BibitemShut {NoStop}%
\bibitem [{\citenamefont {Luo}\ \emph {et~al.}(2017)\citenamefont {Luo},
  \citenamefont {Zou}, \citenamefont {Wu}, \citenamefont {Liu}, \citenamefont
  {Han}, \citenamefont {Tey},\ and\ \citenamefont {You}}]{luo:2017ab}%
  \BibitemOpen
  \bibfield  {author} {\bibinfo {author} {\bibfnamefont {X.-Y.}\ \bibnamefont
  {Luo}}, \bibinfo {author} {\bibfnamefont {Y.-Q.}\ \bibnamefont {Zou}},
  \bibinfo {author} {\bibfnamefont {L.-N.}\ \bibnamefont {Wu}}, \bibinfo
  {author} {\bibfnamefont {Q.}~\bibnamefont {Liu}}, \bibinfo {author}
  {\bibfnamefont {M.-F.}\ \bibnamefont {Han}}, \bibinfo {author} {\bibfnamefont
  {M.~K.}\ \bibnamefont {Tey}},\ and\ \bibinfo {author} {\bibfnamefont
  {L.}~\bibnamefont {You}},\ }\bibfield  {title} {\bibinfo {title}
  {Deterministic entanglement generation from driving through quantum phase
  transitions},\ }\href
  {https://www.science.org/doi/abs/10.1126/science.aag1106} {\bibfield
  {journal} {\bibinfo  {journal} {Science}\ }\textbf {\bibinfo {volume}
  {355}},\ \bibinfo {pages} {620} (\bibinfo {year} {2017})}\BibitemShut
  {NoStop}%
\bibitem [{\citenamefont {Evrard}\ \emph {et~al.}(2021)\citenamefont {Evrard},
  \citenamefont {Qu}, \citenamefont {Dalibard},\ and\ \citenamefont
  {Gerbier}}]{evrard:2021aa}%
  \BibitemOpen
  \bibfield  {author} {\bibinfo {author} {\bibfnamefont {B.}~\bibnamefont
  {Evrard}}, \bibinfo {author} {\bibfnamefont {A.}~\bibnamefont {Qu}}, \bibinfo
  {author} {\bibfnamefont {J.}~\bibnamefont {Dalibard}},\ and\ \bibinfo
  {author} {\bibfnamefont {F.}~\bibnamefont {Gerbier}},\ }\bibfield  {title}
  {\bibinfo {title} {Observation of fragmentation of a spinor bose-einstein
  condensate},\ }\href
  {https://www.science.org/doi/abs/10.1126/science.abd8206} {\bibfield
  {journal} {\bibinfo  {journal} {Science}\ }\textbf {\bibinfo {volume}
  {373}},\ \bibinfo {pages} {1340} (\bibinfo {year} {2021})}\BibitemShut
  {NoStop}%
\bibitem [{\citenamefont {Jo}\ \emph {et~al.}(2009)\citenamefont {Jo},
  \citenamefont {Lee}, \citenamefont {Choi}, \citenamefont {Christensen},
  \citenamefont {Kim}, \citenamefont {Thywissen}, \citenamefont {Pritchard},\
  and\ \citenamefont {Ketterle}}]{JoIFIAFGOUA2009}%
  \BibitemOpen
  \bibfield  {author} {\bibinfo {author} {\bibfnamefont {G.-B.}\ \bibnamefont
  {Jo}}, \bibinfo {author} {\bibfnamefont {Y.-R.}\ \bibnamefont {Lee}},
  \bibinfo {author} {\bibfnamefont {J.-H.}\ \bibnamefont {Choi}}, \bibinfo
  {author} {\bibfnamefont {C.~A.}\ \bibnamefont {Christensen}}, \bibinfo
  {author} {\bibfnamefont {T.~H.}\ \bibnamefont {Kim}}, \bibinfo {author}
  {\bibfnamefont {J.~H.}\ \bibnamefont {Thywissen}}, \bibinfo {author}
  {\bibfnamefont {D.~E.}\ \bibnamefont {Pritchard}},\ and\ \bibinfo {author}
  {\bibfnamefont {W.}~\bibnamefont {Ketterle}},\ }\bibfield  {title} {\bibinfo
  {title} {Itinerant ferromagnetism in a fermi gas of ultracold atoms},\ }\href
  {https://doi.org/10.1126/science.1177112} {\bibfield  {journal} {\bibinfo
  {journal} {Science}\ }\textbf {\bibinfo {volume} {325}},\ \bibinfo {pages}
  {1521} (\bibinfo {year} {2009})}\BibitemShut {NoStop}%
\bibitem [{\citenamefont {Valtolina}\ \emph {et~al.}(2017)\citenamefont
  {Valtolina}, \citenamefont {Scazza}, \citenamefont {Amico}, \citenamefont
  {Burchianti}, \citenamefont {Recati}, \citenamefont {Enss}, \citenamefont
  {Inguscio}, \citenamefont {Zaccanti},\ and\ \citenamefont
  {Roati}}]{Valtolina:2017aa}%
  \BibitemOpen
  \bibfield  {author} {\bibinfo {author} {\bibfnamefont {G.}~\bibnamefont
  {Valtolina}}, \bibinfo {author} {\bibfnamefont {F.}~\bibnamefont {Scazza}},
  \bibinfo {author} {\bibfnamefont {A.}~\bibnamefont {Amico}}, \bibinfo
  {author} {\bibfnamefont {A.}~\bibnamefont {Burchianti}}, \bibinfo {author}
  {\bibfnamefont {A.}~\bibnamefont {Recati}}, \bibinfo {author} {\bibfnamefont
  {T.}~\bibnamefont {Enss}}, \bibinfo {author} {\bibfnamefont {M.}~\bibnamefont
  {Inguscio}}, \bibinfo {author} {\bibfnamefont {M.}~\bibnamefont {Zaccanti}},\
  and\ \bibinfo {author} {\bibfnamefont {G.}~\bibnamefont {Roati}},\ }\bibfield
   {title} {\bibinfo {title} {Exploring the ferromagnetic behaviour of a
  repulsive fermi gas through spin dynamics},\ }\href
  {https://doi.org/10.1038/nphys4108} {\bibfield  {journal} {\bibinfo
  {journal} {Nat Phys}\ }\textbf {\bibinfo {volume} {13}},\ \bibinfo {pages}
  {704} (\bibinfo {year} {2017})}\BibitemShut {NoStop}%
\bibitem [{\citenamefont {Sanner}\ \emph {et~al.}(2011)\citenamefont {Sanner},
  \citenamefont {Su}, \citenamefont {Keshet}, \citenamefont {Huang},
  \citenamefont {Gillen}, \citenamefont {Gommers},\ and\ \citenamefont
  {Ketterle}}]{Sanner:2011aa}%
  \BibitemOpen
  \bibfield  {author} {\bibinfo {author} {\bibfnamefont {C.}~\bibnamefont
  {Sanner}}, \bibinfo {author} {\bibfnamefont {E.~J.}\ \bibnamefont {Su}},
  \bibinfo {author} {\bibfnamefont {A.}~\bibnamefont {Keshet}}, \bibinfo
  {author} {\bibfnamefont {W.}~\bibnamefont {Huang}}, \bibinfo {author}
  {\bibfnamefont {J.}~\bibnamefont {Gillen}}, \bibinfo {author} {\bibfnamefont
  {R.}~\bibnamefont {Gommers}},\ and\ \bibinfo {author} {\bibfnamefont
  {W.}~\bibnamefont {Ketterle}},\ }\bibfield  {title} {\bibinfo {title}
  {Speckle imaging of spin fluctuations in a strongly interacting fermi gas},\
  }\href {https://link.aps.org/doi/10.1103/PhysRevLett.106.010402} {\bibfield
  {journal} {\bibinfo  {journal} {Phys. Rev. Lett.}\ }\textbf {\bibinfo
  {volume} {106}},\ \bibinfo {pages} {010402} (\bibinfo {year}
  {2011})}\BibitemShut {NoStop}%
\bibitem [{\citenamefont {Amico}\ \emph {et~al.}(2018)\citenamefont {Amico},
  \citenamefont {Scazza}, \citenamefont {Valtolina}, \citenamefont {Tavares},
  \citenamefont {Ketterle}, \citenamefont {Inguscio}, \citenamefont {Roati},\
  and\ \citenamefont {Zaccanti}}]{AmicoTROOCAARSRCISIFG2018}%
  \BibitemOpen
  \bibfield  {author} {\bibinfo {author} {\bibfnamefont {A.}~\bibnamefont
  {Amico}}, \bibinfo {author} {\bibfnamefont {F.}~\bibnamefont {Scazza}},
  \bibinfo {author} {\bibfnamefont {G.}~\bibnamefont {Valtolina}}, \bibinfo
  {author} {\bibfnamefont {P.~E.~S.}\ \bibnamefont {Tavares}}, \bibinfo
  {author} {\bibfnamefont {W.}~\bibnamefont {Ketterle}}, \bibinfo {author}
  {\bibfnamefont {M.}~\bibnamefont {Inguscio}}, \bibinfo {author}
  {\bibfnamefont {G.}~\bibnamefont {Roati}},\ and\ \bibinfo {author}
  {\bibfnamefont {M.}~\bibnamefont {Zaccanti}},\ }\bibfield  {title} {\bibinfo
  {title} {Time-resolved observation of competing attractive and repulsive
  short-range correlations in strongly interacting fermi gases},\ }\href
  {https://doi.org/10.1103/PhysRevLett.121.253602} {\bibfield  {journal}
  {\bibinfo  {journal} {Phys. Rev. Lett.}\ }\textbf {\bibinfo {volume} {121}},\
  \bibinfo {pages} {253602} (\bibinfo {year} {2018})}\BibitemShut {NoStop}%
\bibitem [{\citenamefont {De~Marco}\ \emph {et~al.}(2019)\citenamefont
  {De~Marco}, \citenamefont {Valtolina}, \citenamefont {Matsuda}, \citenamefont
  {Tobias}, \citenamefont {Covey},\ and\ \citenamefont {Ye}}]{De-Marco:2019aa}%
  \BibitemOpen
  \bibfield  {author} {\bibinfo {author} {\bibfnamefont {L.}~\bibnamefont
  {De~Marco}}, \bibinfo {author} {\bibfnamefont {G.}~\bibnamefont {Valtolina}},
  \bibinfo {author} {\bibfnamefont {K.}~\bibnamefont {Matsuda}}, \bibinfo
  {author} {\bibfnamefont {W.~G.}\ \bibnamefont {Tobias}}, \bibinfo {author}
  {\bibfnamefont {J.~P.}\ \bibnamefont {Covey}},\ and\ \bibinfo {author}
  {\bibfnamefont {J.}~\bibnamefont {Ye}},\ }\bibfield  {title} {\bibinfo
  {title} {A degenerate fermi gas of polar molecules},\ }\href
  {https://www.science.org/doi/abs/10.1126/science.aau7230} {\bibfield
  {journal} {\bibinfo  {journal} {Science}\ }\textbf {\bibinfo {volume}
  {363}},\ \bibinfo {pages} {853} (\bibinfo {year} {2019})}\BibitemShut
  {NoStop}%
\bibitem [{\citenamefont {Chen}\ \emph {et~al.}(2023)\citenamefont {Chen},
  \citenamefont {Schindewolf}, \citenamefont {Eppelt}, \citenamefont {Bause},
  \citenamefont {Duda}, \citenamefont {Biswas}, \citenamefont {Karman},
  \citenamefont {Hilker}, \citenamefont {Bloch},\ and\ \citenamefont
  {Luo}}]{chen:2023aa}%
  \BibitemOpen
  \bibfield  {author} {\bibinfo {author} {\bibfnamefont {X.-Y.}\ \bibnamefont
  {Chen}}, \bibinfo {author} {\bibfnamefont {A.}~\bibnamefont {Schindewolf}},
  \bibinfo {author} {\bibfnamefont {S.}~\bibnamefont {Eppelt}}, \bibinfo
  {author} {\bibfnamefont {R.}~\bibnamefont {Bause}}, \bibinfo {author}
  {\bibfnamefont {M.}~\bibnamefont {Duda}}, \bibinfo {author} {\bibfnamefont
  {S.}~\bibnamefont {Biswas}}, \bibinfo {author} {\bibfnamefont
  {T.}~\bibnamefont {Karman}}, \bibinfo {author} {\bibfnamefont
  {T.}~\bibnamefont {Hilker}}, \bibinfo {author} {\bibfnamefont
  {I.}~\bibnamefont {Bloch}},\ and\ \bibinfo {author} {\bibfnamefont {X.-Y.}\
  \bibnamefont {Luo}},\ }\bibfield  {title} {\bibinfo {title} {Field-linked
  resonances of polar molecules},\ }\href
  {https://doi.org/10.1038/s41586-022-05651-8} {\bibfield  {journal} {\bibinfo
  {journal} {Nature}\ }\textbf {\bibinfo {volume} {614}},\ \bibinfo {pages}
  {59} (\bibinfo {year} {2023})}\BibitemShut {NoStop}%
\bibitem [{\citenamefont {Zhang}\ \emph {et~al.}(2021)\citenamefont {Zhang},
  \citenamefont {Chen}, \citenamefont {Wu}, \citenamefont {Wang}, \citenamefont
  {Fan}, \citenamefont {Deng},\ and\ \citenamefont {Wu}}]{zhang:2021tr}%
  \BibitemOpen
  \bibfield  {author} {\bibinfo {author} {\bibfnamefont {X.}~\bibnamefont
  {Zhang}}, \bibinfo {author} {\bibfnamefont {Y.}~\bibnamefont {Chen}},
  \bibinfo {author} {\bibfnamefont {Z.}~\bibnamefont {Wu}}, \bibinfo {author}
  {\bibfnamefont {J.}~\bibnamefont {Wang}}, \bibinfo {author} {\bibfnamefont
  {J.}~\bibnamefont {Fan}}, \bibinfo {author} {\bibfnamefont {S.}~\bibnamefont
  {Deng}},\ and\ \bibinfo {author} {\bibfnamefont {H.}~\bibnamefont {Wu}},\
  }\bibfield  {title} {\bibinfo {title} {Observation of a superradiant quantum
  phase transition in an intracavity degenerate fermi gas},\ }\href
  {https://www.science.org/doi/abs/10.1126/science.abd4385} {\bibfield
  {journal} {\bibinfo  {journal} {Science}\ }\textbf {\bibinfo {volume}
  {373}},\ \bibinfo {pages} {1359} (\bibinfo {year} {2021})}\BibitemShut
  {NoStop}%
\bibitem [{\citenamefont {Helson}\ \emph
  {et~al.}(2023{\natexlab{a}})\citenamefont {Helson}, \citenamefont {Zwettler},
  \citenamefont {Mivehvar}, \citenamefont {Colella}, \citenamefont {Roux},
  \citenamefont {Konishi}, \citenamefont {Ritsch},\ and\ \citenamefont
  {Brantut}}]{HelsonDWOIAUFGWPMI2023}%
  \BibitemOpen
  \bibfield  {author} {\bibinfo {author} {\bibfnamefont {V.}~\bibnamefont
  {Helson}}, \bibinfo {author} {\bibfnamefont {T.}~\bibnamefont {Zwettler}},
  \bibinfo {author} {\bibfnamefont {F.}~\bibnamefont {Mivehvar}}, \bibinfo
  {author} {\bibfnamefont {E.}~\bibnamefont {Colella}}, \bibinfo {author}
  {\bibfnamefont {K.}~\bibnamefont {Roux}}, \bibinfo {author} {\bibfnamefont
  {H.}~\bibnamefont {Konishi}}, \bibinfo {author} {\bibfnamefont
  {H.}~\bibnamefont {Ritsch}},\ and\ \bibinfo {author} {\bibfnamefont {J.-P.}\
  \bibnamefont {Brantut}},\ }\bibfield  {title} {\bibinfo {title} {Density-wave
  ordering in a unitary fermi gas with photon-mediated interactions},\ }\href
  {https://doi.org/10.1038/s41586-023-06018-3} {\bibfield  {journal} {\bibinfo
  {journal} {Nature}\ }\textbf {\bibinfo {volume} {618}},\ \bibinfo {pages}
  {716} (\bibinfo {year} {2023}{\natexlab{a}})}\BibitemShut {NoStop}%
\bibitem [{\citenamefont {Cao}\ \emph {et~al.}(2011)\citenamefont {Cao},
  \citenamefont {Elliott}, \citenamefont {Joseph}, \citenamefont {Wu},
  \citenamefont {Petricka}, \citenamefont {Sch{\"a}fer},\ and\ \citenamefont
  {Thomas}}]{Cao:2011ab}%
  \BibitemOpen
  \bibfield  {author} {\bibinfo {author} {\bibfnamefont {C.}~\bibnamefont
  {Cao}}, \bibinfo {author} {\bibfnamefont {E.}~\bibnamefont {Elliott}},
  \bibinfo {author} {\bibfnamefont {J.}~\bibnamefont {Joseph}}, \bibinfo
  {author} {\bibfnamefont {H.}~\bibnamefont {Wu}}, \bibinfo {author}
  {\bibfnamefont {J.}~\bibnamefont {Petricka}}, \bibinfo {author}
  {\bibfnamefont {T.}~\bibnamefont {Sch{\"a}fer}},\ and\ \bibinfo {author}
  {\bibfnamefont {J.~E.}\ \bibnamefont {Thomas}},\ }\bibfield  {title}
  {\bibinfo {title} {Universal quantum viscosity in a unitary fermi gas},\
  }\href {https://www.science.org/doi/abs/10.1126/science.1195219} {\bibfield
  {journal} {\bibinfo  {journal} {Science}\ }\textbf {\bibinfo {volume}
  {331}},\ \bibinfo {pages} {58} (\bibinfo {year} {2011})}\BibitemShut
  {NoStop}%
\bibitem [{\citenamefont {Koschorreck}\ \emph {et~al.}(2013)\citenamefont
  {Koschorreck}, \citenamefont {Pertot}, \citenamefont {Vogt},\ and\
  \citenamefont {Kohl}}]{Koschorreck:2013aa}%
  \BibitemOpen
  \bibfield  {author} {\bibinfo {author} {\bibfnamefont {M.}~\bibnamefont
  {Koschorreck}}, \bibinfo {author} {\bibfnamefont {D.}~\bibnamefont {Pertot}},
  \bibinfo {author} {\bibfnamefont {E.}~\bibnamefont {Vogt}},\ and\ \bibinfo
  {author} {\bibfnamefont {M.}~\bibnamefont {Kohl}},\ }\bibfield  {title}
  {\bibinfo {title} {Universal spin dynamics in two-dimensional fermi gases},\
  }\href {https://doi.org/10.1038/nphys2637} {\bibfield  {journal} {\bibinfo
  {journal} {Nat Phys}\ }\textbf {\bibinfo {volume} {9}},\ \bibinfo {pages}
  {405} (\bibinfo {year} {2013})}\BibitemShut {NoStop}%
\bibitem [{\citenamefont {Bardon}\ \emph {et~al.}(2014)\citenamefont {Bardon},
  \citenamefont {Beattie}, \citenamefont {Luciuk}, \citenamefont {Cairncross},
  \citenamefont {Fine}, \citenamefont {Cheng}, \citenamefont {Edge},
  \citenamefont {Taylor}, \citenamefont {Zhang}, \citenamefont {Trotzky},\ and\
  \citenamefont {Thywissen}}]{Bardon:2014aa}%
  \BibitemOpen
  \bibfield  {author} {\bibinfo {author} {\bibfnamefont {A.~B.}\ \bibnamefont
  {Bardon}}, \bibinfo {author} {\bibfnamefont {S.}~\bibnamefont {Beattie}},
  \bibinfo {author} {\bibfnamefont {C.}~\bibnamefont {Luciuk}}, \bibinfo
  {author} {\bibfnamefont {W.}~\bibnamefont {Cairncross}}, \bibinfo {author}
  {\bibfnamefont {D.}~\bibnamefont {Fine}}, \bibinfo {author} {\bibfnamefont
  {N.~S.}\ \bibnamefont {Cheng}}, \bibinfo {author} {\bibfnamefont {G.~J.~A.}\
  \bibnamefont {Edge}}, \bibinfo {author} {\bibfnamefont {E.}~\bibnamefont
  {Taylor}}, \bibinfo {author} {\bibfnamefont {S.}~\bibnamefont {Zhang}},
  \bibinfo {author} {\bibfnamefont {S.}~\bibnamefont {Trotzky}},\ and\ \bibinfo
  {author} {\bibfnamefont {J.~H.}\ \bibnamefont {Thywissen}},\ }\bibfield
  {title} {\bibinfo {title} {Transverse demagnetization dynamics of a unitary
  fermi gas},\ }\href {https://www.science.org/doi/abs/10.1126/science.1247425}
  {\bibfield  {journal} {\bibinfo  {journal} {Science}\ }\textbf {\bibinfo
  {volume} {344}},\ \bibinfo {pages} {722} (\bibinfo {year}
  {2014})}\BibitemShut {NoStop}%
\bibitem [{\citenamefont {Joseph}\ \emph {et~al.}(2015)\citenamefont {Joseph},
  \citenamefont {Elliott},\ and\ \citenamefont {Thomas}}]{Joseph:2015aa}%
  \BibitemOpen
  \bibfield  {author} {\bibinfo {author} {\bibfnamefont {J.~A.}\ \bibnamefont
  {Joseph}}, \bibinfo {author} {\bibfnamefont {E.}~\bibnamefont {Elliott}},\
  and\ \bibinfo {author} {\bibfnamefont {J.~E.}\ \bibnamefont {Thomas}},\
  }\bibfield  {title} {\bibinfo {title} {Shear viscosity of a unitary fermi gas
  near the superfluid phase transition},\ }\href
  {https://link.aps.org/doi/10.1103/PhysRevLett.115.020401} {\bibfield
  {journal} {\bibinfo  {journal} {Phys. Rev. Lett.}\ }\textbf {\bibinfo
  {volume} {115}},\ \bibinfo {pages} {020401} (\bibinfo {year}
  {2015})}\BibitemShut {NoStop}%
\bibitem [{\citenamefont {Luciuk}\ \emph {et~al.}(2017)\citenamefont {Luciuk},
  \citenamefont {Smale}, \citenamefont {B\"ottcher}, \citenamefont {Sharum},
  \citenamefont {Olsen}, \citenamefont {Trotzky}, \citenamefont {Enss},\ and\
  \citenamefont {Thywissen}}]{Luciuk:2017aa}%
  \BibitemOpen
  \bibfield  {author} {\bibinfo {author} {\bibfnamefont {C.}~\bibnamefont
  {Luciuk}}, \bibinfo {author} {\bibfnamefont {S.}~\bibnamefont {Smale}},
  \bibinfo {author} {\bibfnamefont {F.}~\bibnamefont {B\"ottcher}}, \bibinfo
  {author} {\bibfnamefont {H.}~\bibnamefont {Sharum}}, \bibinfo {author}
  {\bibfnamefont {B.~A.}\ \bibnamefont {Olsen}}, \bibinfo {author}
  {\bibfnamefont {S.}~\bibnamefont {Trotzky}}, \bibinfo {author} {\bibfnamefont
  {T.}~\bibnamefont {Enss}},\ and\ \bibinfo {author} {\bibfnamefont {J.~H.}\
  \bibnamefont {Thywissen}},\ }\bibfield  {title} {\bibinfo {title}
  {Observation of quantum-limited spin transport in strongly interacting
  two-dimensional fermi gases},\ }\href
  {https://link.aps.org/doi/10.1103/PhysRevLett.118.130405} {\bibfield
  {journal} {\bibinfo  {journal} {Phys. Rev. Lett.}\ }\textbf {\bibinfo
  {volume} {118}},\ \bibinfo {pages} {130405} (\bibinfo {year}
  {2017})}\BibitemShut {NoStop}%
\bibitem [{\citenamefont {Patel}\ \emph {et~al.}(2020)\citenamefont {Patel},
  \citenamefont {Yan}, \citenamefont {Mukherjee}, \citenamefont {Fletcher},
  \citenamefont {Struck},\ and\ \citenamefont {Zwierlein}}]{patel:2020th}%
  \BibitemOpen
  \bibfield  {author} {\bibinfo {author} {\bibfnamefont {P.~B.}\ \bibnamefont
  {Patel}}, \bibinfo {author} {\bibfnamefont {Z.}~\bibnamefont {Yan}}, \bibinfo
  {author} {\bibfnamefont {B.}~\bibnamefont {Mukherjee}}, \bibinfo {author}
  {\bibfnamefont {R.~J.}\ \bibnamefont {Fletcher}}, \bibinfo {author}
  {\bibfnamefont {J.}~\bibnamefont {Struck}},\ and\ \bibinfo {author}
  {\bibfnamefont {M.~W.}\ \bibnamefont {Zwierlein}},\ }\bibfield  {title}
  {\bibinfo {title} {Universal sound diffusion in a strongly interacting fermi
  gas},\ }\href {https://www.science.org/doi/abs/10.1126/science.aaz5756}
  {\bibfield  {journal} {\bibinfo  {journal} {Science}\ }\textbf {\bibinfo
  {volume} {370}},\ \bibinfo {pages} {1222} (\bibinfo {year}
  {2020})}\BibitemShut {NoStop}%
\bibitem [{\citenamefont {Defenu}\ \emph {et~al.}(2023)\citenamefont {Defenu},
  \citenamefont {Donner}, \citenamefont {Macr\`{\i}}, \citenamefont {Pagano},
  \citenamefont {Ruffo},\ and\ \citenamefont {Trombettoni}}]{defenu:2023aa}%
  \BibitemOpen
  \bibfield  {author} {\bibinfo {author} {\bibfnamefont {N.}~\bibnamefont
  {Defenu}}, \bibinfo {author} {\bibfnamefont {T.}~\bibnamefont {Donner}},
  \bibinfo {author} {\bibfnamefont {T.}~\bibnamefont {Macr\`{\i}}}, \bibinfo
  {author} {\bibfnamefont {G.}~\bibnamefont {Pagano}}, \bibinfo {author}
  {\bibfnamefont {S.}~\bibnamefont {Ruffo}},\ and\ \bibinfo {author}
  {\bibfnamefont {A.}~\bibnamefont {Trombettoni}},\ }\bibfield  {title}
  {\bibinfo {title} {Long-range interacting quantum systems},\ }\href
  {https://link.aps.org/doi/10.1103/RevModPhys.95.035002} {\bibfield  {journal}
  {\bibinfo  {journal} {Rev. Mod. Phys.}\ }\textbf {\bibinfo {volume} {95}},\
  \bibinfo {pages} {035002} (\bibinfo {year} {2023})}\BibitemShut {NoStop}%
\bibitem [{\citenamefont {{Baumann}}\ \emph {et~al.}(2010)\citenamefont
  {{Baumann}}, \citenamefont {{Guerlin}}, \citenamefont {{Brennecke}},\ and\
  \citenamefont {{Esslinger}}}]{Baumann:2010aa}%
  \BibitemOpen
  \bibfield  {author} {\bibinfo {author} {\bibfnamefont {K.}~\bibnamefont
  {{Baumann}}}, \bibinfo {author} {\bibfnamefont {C.}~\bibnamefont
  {{Guerlin}}}, \bibinfo {author} {\bibfnamefont {F.}~\bibnamefont
  {{Brennecke}}},\ and\ \bibinfo {author} {\bibfnamefont {T.}~\bibnamefont
  {{Esslinger}}},\ }\bibfield  {title} {\bibinfo {title} {Dicke quantum phase
  transition with a superfluid gas in an optical cavity},\ }\href
  {https://doi.org/10.1038/nature09009} {\bibfield  {journal} {\bibinfo
  {journal} {Nature}\ }\textbf {\bibinfo {volume} {464}},\ \bibinfo {pages}
  {1301} (\bibinfo {year} {2010})}\BibitemShut {NoStop}%
\bibitem [{\citenamefont {Klinder}\ \emph {et~al.}(2015)\citenamefont
  {Klinder}, \citenamefont {Keßler}, \citenamefont {Wolke}, \citenamefont
  {Mathey},\ and\ \citenamefont {Hemmerich}}]{KlinderDPTITODM2015}%
  \BibitemOpen
  \bibfield  {author} {\bibinfo {author} {\bibfnamefont {J.}~\bibnamefont
  {Klinder}}, \bibinfo {author} {\bibfnamefont {H.}~\bibnamefont {Keßler}},
  \bibinfo {author} {\bibfnamefont {M.}~\bibnamefont {Wolke}}, \bibinfo
  {author} {\bibfnamefont {L.}~\bibnamefont {Mathey}},\ and\ \bibinfo {author}
  {\bibfnamefont {A.}~\bibnamefont {Hemmerich}},\ }\bibfield  {title} {\bibinfo
  {title} {Dynamical phase transition in the open dicke model},\ }\href
  {https://doi.org/10.1073/pnas.1417132112} {\bibfield  {journal} {\bibinfo
  {journal} {Proceedings of the National Academy of Sciences}\ }\textbf
  {\bibinfo {volume} {112}},\ \bibinfo {pages} {3290} (\bibinfo {year}
  {2015})}\BibitemShut {NoStop}%
\bibitem [{\citenamefont {Ritsch}\ \emph {et~al.}(2013)\citenamefont {Ritsch},
  \citenamefont {Domokos}, \citenamefont {Brennecke},\ and\ \citenamefont
  {Esslinger}}]{Ritsch:2013aa}%
  \BibitemOpen
  \bibfield  {author} {\bibinfo {author} {\bibfnamefont {H.}~\bibnamefont
  {Ritsch}}, \bibinfo {author} {\bibfnamefont {P.}~\bibnamefont {Domokos}},
  \bibinfo {author} {\bibfnamefont {F.}~\bibnamefont {Brennecke}},\ and\
  \bibinfo {author} {\bibfnamefont {T.}~\bibnamefont {Esslinger}},\ }\bibfield
  {title} {\bibinfo {title} {Cold atoms in cavity-generated dynamical optical
  potentials},\ }\href {https://link.aps.org/doi/10.1103/RevModPhys.85.553}
  {\bibfield  {journal} {\bibinfo  {journal} {Rev. Mod. Phys.}\ }\textbf
  {\bibinfo {volume} {85}},\ \bibinfo {pages} {553} (\bibinfo {year}
  {2013})}\BibitemShut {NoStop}%
\bibitem [{\citenamefont {Vaidya}\ \emph {et~al.}(2018)\citenamefont {Vaidya},
  \citenamefont {Guo}, \citenamefont {Kroeze}, \citenamefont {Ballantine},
  \citenamefont {Koll\'ar}, \citenamefont {Keeling},\ and\ \citenamefont
  {Lev}}]{Vaidya:2018aa}%
  \BibitemOpen
  \bibfield  {author} {\bibinfo {author} {\bibfnamefont {V.~D.}\ \bibnamefont
  {Vaidya}}, \bibinfo {author} {\bibfnamefont {Y.}~\bibnamefont {Guo}},
  \bibinfo {author} {\bibfnamefont {R.~M.}\ \bibnamefont {Kroeze}}, \bibinfo
  {author} {\bibfnamefont {K.~E.}\ \bibnamefont {Ballantine}}, \bibinfo
  {author} {\bibfnamefont {A.~J.}\ \bibnamefont {Koll\'ar}}, \bibinfo {author}
  {\bibfnamefont {J.}~\bibnamefont {Keeling}},\ and\ \bibinfo {author}
  {\bibfnamefont {B.~L.}\ \bibnamefont {Lev}},\ }\bibfield  {title} {\bibinfo
  {title} {Tunable-range, photon-mediated atomic interactions in multimode
  cavity qed},\ }\href {https://link.aps.org/doi/10.1103/PhysRevX.8.011002}
  {\bibfield  {journal} {\bibinfo  {journal} {Phys. Rev. X}\ }\textbf {\bibinfo
  {volume} {8}},\ \bibinfo {pages} {011002} (\bibinfo {year}
  {2018})}\BibitemShut {NoStop}%
\bibitem [{\citenamefont {Mivehvar}\ \emph {et~al.}(2021)\citenamefont
  {Mivehvar}, \citenamefont {Piazza}, \citenamefont {Donner},\ and\
  \citenamefont {Ritsch}}]{mivehvar:2021aa}%
  \BibitemOpen
  \bibfield  {author} {\bibinfo {author} {\bibfnamefont {F.}~\bibnamefont
  {Mivehvar}}, \bibinfo {author} {\bibfnamefont {F.}~\bibnamefont {Piazza}},
  \bibinfo {author} {\bibfnamefont {T.}~\bibnamefont {Donner}},\ and\ \bibinfo
  {author} {\bibfnamefont {H.}~\bibnamefont {Ritsch}},\ }\bibfield  {title}
  {\bibinfo {title} {Cavity qed with quantum gases: new paradigms in many-body
  physics},\ }\href {https://doi.org/10.1080/00018732.2021.1969727} {\bibfield
  {journal} {\bibinfo  {journal} {Advances in Physics}\ }\textbf {\bibinfo
  {volume} {70}},\ \bibinfo {pages} {1} (\bibinfo {year} {2021})}\BibitemShut
  {NoStop}%
\bibitem [{\citenamefont {Cetina}\ \emph {et~al.}(2016)\citenamefont {Cetina},
  \citenamefont {Jag}, \citenamefont {Lous}, \citenamefont {Fritsche},
  \citenamefont {Walraven}, \citenamefont {Grimm}, \citenamefont {Levinsen},
  \citenamefont {Parish}, \citenamefont {Schmidt}, \citenamefont {Knap},\ and\
  \citenamefont {Demler}}]{CetinaUMBIOICTAFS2016}%
  \BibitemOpen
  \bibfield  {author} {\bibinfo {author} {\bibfnamefont {M.}~\bibnamefont
  {Cetina}}, \bibinfo {author} {\bibfnamefont {M.}~\bibnamefont {Jag}},
  \bibinfo {author} {\bibfnamefont {R.~S.}\ \bibnamefont {Lous}}, \bibinfo
  {author} {\bibfnamefont {I.}~\bibnamefont {Fritsche}}, \bibinfo {author}
  {\bibfnamefont {J.~T.~M.}\ \bibnamefont {Walraven}}, \bibinfo {author}
  {\bibfnamefont {R.}~\bibnamefont {Grimm}}, \bibinfo {author} {\bibfnamefont
  {J.}~\bibnamefont {Levinsen}}, \bibinfo {author} {\bibfnamefont {M.~M.}\
  \bibnamefont {Parish}}, \bibinfo {author} {\bibfnamefont {R.}~\bibnamefont
  {Schmidt}}, \bibinfo {author} {\bibfnamefont {M.}~\bibnamefont {Knap}},\ and\
  \bibinfo {author} {\bibfnamefont {E.}~\bibnamefont {Demler}},\ }\bibfield
  {title} {\bibinfo {title} {Ultrafast many-body interferometry of impurities
  coupled to a fermi sea},\ }\href {https://doi.org/10.1126/science.aaf5134}
  {\bibfield  {journal} {\bibinfo  {journal} {Science}\ }\textbf {\bibinfo
  {volume} {354}},\ \bibinfo {pages} {96} (\bibinfo {year} {2016})}\BibitemShut
  {NoStop}%
\bibitem [{\citenamefont {Pekker}\ \emph {et~al.}(2011)\citenamefont {Pekker},
  \citenamefont {Babadi}, \citenamefont {Sensarma}, \citenamefont {Zinner},
  \citenamefont {Pollet}, \citenamefont {Zwierlein},\ and\ \citenamefont
  {Demler}}]{PekkerCBPAFIUFGNFR2011}%
  \BibitemOpen
  \bibfield  {author} {\bibinfo {author} {\bibfnamefont {D.}~\bibnamefont
  {Pekker}}, \bibinfo {author} {\bibfnamefont {M.}~\bibnamefont {Babadi}},
  \bibinfo {author} {\bibfnamefont {R.}~\bibnamefont {Sensarma}}, \bibinfo
  {author} {\bibfnamefont {N.}~\bibnamefont {Zinner}}, \bibinfo {author}
  {\bibfnamefont {L.}~\bibnamefont {Pollet}}, \bibinfo {author} {\bibfnamefont
  {M.~W.}\ \bibnamefont {Zwierlein}},\ and\ \bibinfo {author} {\bibfnamefont
  {E.}~\bibnamefont {Demler}},\ }\bibfield  {title} {\bibinfo {title}
  {Competition between pairing and ferromagnetic instabilities in ultracold
  fermi gases near feshbach resonances},\ }\href
  {https://doi.org/10.1103/PhysRevLett.106.050402} {\bibfield  {journal}
  {\bibinfo  {journal} {Phys. Rev. Lett.}\ }\textbf {\bibinfo {volume} {106}},\
  \bibinfo {pages} {050402} (\bibinfo {year} {2011})}\BibitemShut {NoStop}%
\bibitem [{\citenamefont {Subramanyan}\ \emph {et~al.}(2021)\citenamefont
  {Subramanyan}, \citenamefont {Hegde}, \citenamefont {Vishveshwara},\ and\
  \citenamefont {Bradlyn}}]{SubramanyanPOTIHO2021}%
  \BibitemOpen
  \bibfield  {author} {\bibinfo {author} {\bibfnamefont {V.}~\bibnamefont
  {Subramanyan}}, \bibinfo {author} {\bibfnamefont {S.~S.}\ \bibnamefont
  {Hegde}}, \bibinfo {author} {\bibfnamefont {S.}~\bibnamefont
  {Vishveshwara}},\ and\ \bibinfo {author} {\bibfnamefont {B.}~\bibnamefont
  {Bradlyn}},\ }\bibfield  {title} {\bibinfo {title} {Physics of the inverted
  harmonic oscillator: From the lowest landau level to event horizons},\ }\href
  {https://doi.org/https://doi.org/10.1016/j.aop.2021.168470} {\bibfield
  {journal} {\bibinfo  {journal} {Annals of Physics}\ }\textbf {\bibinfo
  {volume} {435}},\ \bibinfo {pages} {168470} (\bibinfo {year} {2021})},\
  \bibinfo {note} {special issue on Philip W. Anderson}\BibitemShut {NoStop}%
\bibitem [{\citenamefont {Roux}\ \emph {et~al.}(2020)\citenamefont {Roux},
  \citenamefont {Konishi}, \citenamefont {Helson},\ and\ \citenamefont
  {Brantut}}]{Roux:2020aa}%
  \BibitemOpen
  \bibfield  {author} {\bibinfo {author} {\bibfnamefont {K.}~\bibnamefont
  {Roux}}, \bibinfo {author} {\bibfnamefont {H.}~\bibnamefont {Konishi}},
  \bibinfo {author} {\bibfnamefont {V.}~\bibnamefont {Helson}},\ and\ \bibinfo
  {author} {\bibfnamefont {J.-P.}\ \bibnamefont {Brantut}},\ }\bibfield
  {title} {\bibinfo {title} {Strongly correlated fermions strongly coupled to
  light},\ }\href {https://doi.org/10.1038/s41467-020-16767-8} {\bibfield
  {journal} {\bibinfo  {journal} {Nature Communications}\ }\textbf {\bibinfo
  {volume} {11}},\ \bibinfo {pages} {2974} (\bibinfo {year}
  {2020})}\BibitemShut {NoStop}%
\bibitem [{\citenamefont {Roux}\ \emph {et~al.}(2021)\citenamefont {Roux},
  \citenamefont {Helson}, \citenamefont {Konishi},\ and\ \citenamefont
  {Brantut}}]{roux:2021uf}%
  \BibitemOpen
  \bibfield  {author} {\bibinfo {author} {\bibfnamefont {K.}~\bibnamefont
  {Roux}}, \bibinfo {author} {\bibfnamefont {V.}~\bibnamefont {Helson}},
  \bibinfo {author} {\bibfnamefont {H.}~\bibnamefont {Konishi}},\ and\ \bibinfo
  {author} {\bibfnamefont {J.-P.}\ \bibnamefont {Brantut}},\ }\bibfield
  {title} {\bibinfo {title} {Cavity-assisted preparation and detection of a
  unitary fermi gas},\ }\href {https://dx.doi.org/10.1088/1367-2630/abeb91}
  {\bibfield  {journal} {\bibinfo  {journal} {New Journal of Physics}\ }\textbf
  {\bibinfo {volume} {23}},\ \bibinfo {pages} {043029} (\bibinfo {year}
  {2021})}\BibitemShut {NoStop}%
\bibitem [{\citenamefont {Wu}\ \emph {et~al.}(2023)\citenamefont {Wu},
  \citenamefont {Fan}, \citenamefont {Zhang}, \citenamefont {Qi},\ and\
  \citenamefont {Wu}}]{wu:2023aa}%
  \BibitemOpen
  \bibfield  {author} {\bibinfo {author} {\bibfnamefont {Z.}~\bibnamefont
  {Wu}}, \bibinfo {author} {\bibfnamefont {J.}~\bibnamefont {Fan}}, \bibinfo
  {author} {\bibfnamefont {X.}~\bibnamefont {Zhang}}, \bibinfo {author}
  {\bibfnamefont {J.}~\bibnamefont {Qi}},\ and\ \bibinfo {author}
  {\bibfnamefont {H.}~\bibnamefont {Wu}},\ }\bibfield  {title} {\bibinfo
  {title} {Signatures of prethermalization in a quenched cavity-mediated
  long-range interacting fermi gas},\ }\href
  {https://link.aps.org/doi/10.1103/PhysRevLett.131.243401} {\bibfield
  {journal} {\bibinfo  {journal} {Phys. Rev. Lett.}\ }\textbf {\bibinfo
  {volume} {131}},\ \bibinfo {pages} {243401} (\bibinfo {year}
  {2023})}\BibitemShut {NoStop}%
\bibitem [{\citenamefont {He}(2016)}]{he:2016aa}%
  \BibitemOpen
  \bibfield  {author} {\bibinfo {author} {\bibfnamefont {L.}~\bibnamefont
  {He}},\ }\bibfield  {title} {\bibinfo {title} {Dynamic density and spin
  responses of a superfluid fermi gas in the bcs--bec crossover: Path integral
  formulation and pair fluctuation theory},\ }\href
  {https://www.sciencedirect.com/science/article/pii/S0003491616301312}
  {\bibfield  {journal} {\bibinfo  {journal} {Annals of Physics}\ }\textbf
  {\bibinfo {volume} {373}},\ \bibinfo {pages} {470} (\bibinfo {year}
  {2016})}\BibitemShut {NoStop}%
\bibitem [{\citenamefont {Helson}\ \emph
  {et~al.}(2023{\natexlab{b}})\citenamefont {Helson}, \citenamefont {Zwettler},
  \citenamefont {Mivehvar}, \citenamefont {Colella}, \citenamefont {Roux},
  \citenamefont {Konishi}, \citenamefont {Ritsch},\ and\ \citenamefont
  {Brantut}}]{helson:2023aa}%
  \BibitemOpen
  \bibfield  {author} {\bibinfo {author} {\bibfnamefont {V.}~\bibnamefont
  {Helson}}, \bibinfo {author} {\bibfnamefont {T.}~\bibnamefont {Zwettler}},
  \bibinfo {author} {\bibfnamefont {F.}~\bibnamefont {Mivehvar}}, \bibinfo
  {author} {\bibfnamefont {E.}~\bibnamefont {Colella}}, \bibinfo {author}
  {\bibfnamefont {K.}~\bibnamefont {Roux}}, \bibinfo {author} {\bibfnamefont
  {H.}~\bibnamefont {Konishi}}, \bibinfo {author} {\bibfnamefont
  {H.}~\bibnamefont {Ritsch}},\ and\ \bibinfo {author} {\bibfnamefont {J.-P.}\
  \bibnamefont {Brantut}},\ }\bibfield  {title} {\bibinfo {title} {Density-wave
  ordering in a unitary fermi gas with photon-mediated interactions},\ }\href
  {https://doi.org/10.1038/s41586-023-06018-3} {\bibfield  {journal} {\bibinfo
  {journal} {Nature}\ }\textbf {\bibinfo {volume} {618}},\ \bibinfo {pages}
  {716} (\bibinfo {year} {2023}{\natexlab{b}})}\BibitemShut {NoStop}%
\bibitem [{\citenamefont {Baumann}\ \emph {et~al.}(2011)\citenamefont
  {Baumann}, \citenamefont {Mottl}, \citenamefont {Brennecke},\ and\
  \citenamefont {Esslinger}}]{Baumann:2011vy}%
  \BibitemOpen
  \bibfield  {author} {\bibinfo {author} {\bibfnamefont {K.}~\bibnamefont
  {Baumann}}, \bibinfo {author} {\bibfnamefont {R.}~\bibnamefont {Mottl}},
  \bibinfo {author} {\bibfnamefont {F.}~\bibnamefont {Brennecke}},\ and\
  \bibinfo {author} {\bibfnamefont {T.}~\bibnamefont {Esslinger}},\ }\bibfield
  {title} {\bibinfo {title} {Exploring symmetry breaking at the dicke quantum
  phase transition},\ }\href
  {https://link.aps.org/doi/10.1103/PhysRevLett.107.140402} {\bibfield
  {journal} {\bibinfo  {journal} {Phys. Rev. Lett.}\ }\textbf {\bibinfo
  {volume} {107}},\ \bibinfo {pages} {140402} (\bibinfo {year}
  {2011})}\BibitemShut {NoStop}%
\bibitem [{\citenamefont {Lamacraft}(2007)}]{lamacraft:2007aa}%
  \BibitemOpen
  \bibfield  {author} {\bibinfo {author} {\bibfnamefont {A.}~\bibnamefont
  {Lamacraft}},\ }\bibfield  {title} {\bibinfo {title} {Quantum quenches in a
  spinor condensate},\ }\href
  {https://link.aps.org/doi/10.1103/PhysRevLett.98.160404} {\bibfield
  {journal} {\bibinfo  {journal} {Phys. Rev. Lett.}\ }\textbf {\bibinfo
  {volume} {98}},\ \bibinfo {pages} {160404} (\bibinfo {year}
  {2007})}\BibitemShut {NoStop}%
\bibitem [{\citenamefont {Bhaseen}\ \emph {et~al.}(2012)\citenamefont
  {Bhaseen}, \citenamefont {Mayoh}, \citenamefont {Simons},\ and\ \citenamefont
  {Keeling}}]{Bhaseen:2012aa}%
  \BibitemOpen
  \bibfield  {author} {\bibinfo {author} {\bibfnamefont {M.~J.}\ \bibnamefont
  {Bhaseen}}, \bibinfo {author} {\bibfnamefont {J.}~\bibnamefont {Mayoh}},
  \bibinfo {author} {\bibfnamefont {B.~D.}\ \bibnamefont {Simons}},\ and\
  \bibinfo {author} {\bibfnamefont {J.}~\bibnamefont {Keeling}},\ }\bibfield
  {title} {\bibinfo {title} {Dynamics of nonequilibrium dicke models},\ }\href
  {https://link.aps.org/doi/10.1103/PhysRevA.85.013817} {\bibfield  {journal}
  {\bibinfo  {journal} {Phys. Rev. A}\ }\textbf {\bibinfo {volume} {85}},\
  \bibinfo {pages} {013817} (\bibinfo {year} {2012})}\BibitemShut {NoStop}%
\bibitem [{\citenamefont {Acevedo}\ \emph {et~al.}(2014)\citenamefont
  {Acevedo}, \citenamefont {Quiroga}, \citenamefont {Rodr\'{\i}guez},\ and\
  \citenamefont {Johnson}}]{acevedo:2014aa}%
  \BibitemOpen
  \bibfield  {author} {\bibinfo {author} {\bibfnamefont {O.~L.}\ \bibnamefont
  {Acevedo}}, \bibinfo {author} {\bibfnamefont {L.}~\bibnamefont {Quiroga}},
  \bibinfo {author} {\bibfnamefont {F.~J.}\ \bibnamefont {Rodr\'{\i}guez}},\
  and\ \bibinfo {author} {\bibfnamefont {N.~F.}\ \bibnamefont {Johnson}},\
  }\bibfield  {title} {\bibinfo {title} {New dynamical scaling universality for
  quantum networks across adiabatic quantum phase transitions},\ }\href
  {https://link.aps.org/doi/10.1103/PhysRevLett.112.030403} {\bibfield
  {journal} {\bibinfo  {journal} {Phys. Rev. Lett.}\ }\textbf {\bibinfo
  {volume} {112}},\ \bibinfo {pages} {030403} (\bibinfo {year}
  {2014})}\BibitemShut {NoStop}%
\bibitem [{\citenamefont {Defenu}\ \emph {et~al.}(2018)\citenamefont {Defenu},
  \citenamefont {Enss}, \citenamefont {Kastner},\ and\ \citenamefont
  {Morigi}}]{defenu:2018aa}%
  \BibitemOpen
  \bibfield  {author} {\bibinfo {author} {\bibfnamefont {N.}~\bibnamefont
  {Defenu}}, \bibinfo {author} {\bibfnamefont {T.}~\bibnamefont {Enss}},
  \bibinfo {author} {\bibfnamefont {M.}~\bibnamefont {Kastner}},\ and\ \bibinfo
  {author} {\bibfnamefont {G.}~\bibnamefont {Morigi}},\ }\bibfield  {title}
  {\bibinfo {title} {Dynamical critical scaling of long-range interacting
  quantum magnets},\ }\href
  {https://link.aps.org/doi/10.1103/PhysRevLett.121.240403} {\bibfield
  {journal} {\bibinfo  {journal} {Phys. Rev. Lett.}\ }\textbf {\bibinfo
  {volume} {121}},\ \bibinfo {pages} {240403} (\bibinfo {year}
  {2018})}\BibitemShut {NoStop}%
\bibitem [{\citenamefont {Sch\"utz}\ and\ \citenamefont
  {Morigi}(2014)}]{schutz:2014aa}%
  \BibitemOpen
  \bibfield  {author} {\bibinfo {author} {\bibfnamefont {S.}~\bibnamefont
  {Sch\"utz}}\ and\ \bibinfo {author} {\bibfnamefont {G.}~\bibnamefont
  {Morigi}},\ }\bibfield  {title} {\bibinfo {title} {Prethermalization of atoms
  due to photon-mediated long-range interactions},\ }\href
  {https://link.aps.org/doi/10.1103/PhysRevLett.113.203002} {\bibfield
  {journal} {\bibinfo  {journal} {Phys. Rev. Lett.}\ }\textbf {\bibinfo
  {volume} {113}},\ \bibinfo {pages} {203002} (\bibinfo {year}
  {2014})}\BibitemShut {NoStop}%
\bibitem [{\citenamefont {Sch\"utz}\ \emph {et~al.}(2016)\citenamefont
  {Sch\"utz}, \citenamefont {J\"ager},\ and\ \citenamefont
  {Morigi}}]{schutz:2016aa}%
  \BibitemOpen
  \bibfield  {author} {\bibinfo {author} {\bibfnamefont {S.}~\bibnamefont
  {Sch\"utz}}, \bibinfo {author} {\bibfnamefont {S.~B.}\ \bibnamefont
  {J\"ager}},\ and\ \bibinfo {author} {\bibfnamefont {G.}~\bibnamefont
  {Morigi}},\ }\bibfield  {title} {\bibinfo {title} {Dissipation-assisted
  prethermalization in long-range interacting atomic ensembles},\ }\href
  {https://link.aps.org/doi/10.1103/PhysRevLett.117.083001} {\bibfield
  {journal} {\bibinfo  {journal} {Phys. Rev. Lett.}\ }\textbf {\bibinfo
  {volume} {117}},\ \bibinfo {pages} {083001} (\bibinfo {year}
  {2016})}\BibitemShut {NoStop}%
\bibitem [{\citenamefont {Gadway}\ \emph {et~al.}(2009)\citenamefont {Gadway},
  \citenamefont {Pertot}, \citenamefont {Reimann}, \citenamefont {Cohen},\ and\
  \citenamefont {Schneble}}]{GadwayAOKDDPBRNR2009}%
  \BibitemOpen
  \bibfield  {author} {\bibinfo {author} {\bibfnamefont {B.}~\bibnamefont
  {Gadway}}, \bibinfo {author} {\bibfnamefont {D.}~\bibnamefont {Pertot}},
  \bibinfo {author} {\bibfnamefont {R.}~\bibnamefont {Reimann}}, \bibinfo
  {author} {\bibfnamefont {M.~G.}\ \bibnamefont {Cohen}},\ and\ \bibinfo
  {author} {\bibfnamefont {D.}~\bibnamefont {Schneble}},\ }\bibfield  {title}
  {\bibinfo {title} {Analysis of kapitza-dirac diffraction patterns beyond the
  raman-nath regime},\ }\href {https://doi.org/10.1364/OE.17.019173} {\bibfield
   {journal} {\bibinfo  {journal} {Opt. Express}\ }\textbf {\bibinfo {volume}
  {17}},\ \bibinfo {pages} {19173} (\bibinfo {year} {2009})}\BibitemShut
  {NoStop}%
\bibitem [{\citenamefont {Helson}\ \emph {et~al.}(2022)\citenamefont {Helson},
  \citenamefont {Zwettler}, \citenamefont {Roux}, \citenamefont {Konishi},
  \citenamefont {Uchino},\ and\ \citenamefont {Brantut}}]{HelsonOROASIFG2022}%
  \BibitemOpen
  \bibfield  {author} {\bibinfo {author} {\bibfnamefont {V.}~\bibnamefont
  {Helson}}, \bibinfo {author} {\bibfnamefont {T.}~\bibnamefont {Zwettler}},
  \bibinfo {author} {\bibfnamefont {K.}~\bibnamefont {Roux}}, \bibinfo {author}
  {\bibfnamefont {H.}~\bibnamefont {Konishi}}, \bibinfo {author} {\bibfnamefont
  {S.}~\bibnamefont {Uchino}},\ and\ \bibinfo {author} {\bibfnamefont {J.-P.}\
  \bibnamefont {Brantut}},\ }\bibfield  {title} {\bibinfo {title}
  {Optomechanical response of a strongly interacting fermi gas},\ }\href
  {https://doi.org/10.1103/PhysRevResearch.4.033199} {\bibfield  {journal}
  {\bibinfo  {journal} {Phys. Rev. Research}\ }\textbf {\bibinfo {volume}
  {4}},\ \bibinfo {pages} {033199} (\bibinfo {year} {2022})}\BibitemShut
  {NoStop}%
\bibitem [{\citenamefont {Dupuis}(2023)}]{Dupuis}%
  \BibitemOpen
  \bibfield  {author} {\bibinfo {author} {\bibfnamefont {N.}~\bibnamefont
  {Dupuis}},\ }\href@noop {} {\emph {\bibinfo {title} {Field Theory Of
  Condensed Matter And Ultracold Gases - Volume 1}}},\ \bibinfo {number} {v.
  1}\ (\bibinfo  {publisher} {World Scientific Publishing Europe Limited},\
  \bibinfo {year} {2023})\BibitemShut {NoStop}%
\bibitem [{\citenamefont {Combescot}\ \emph {et~al.}(2006)\citenamefont
  {Combescot}, \citenamefont {Kagan},\ and\ \citenamefont
  {Stringari}}]{Response-sound}%
  \BibitemOpen
  \bibfield  {author} {\bibinfo {author} {\bibfnamefont {R.}~\bibnamefont
  {Combescot}}, \bibinfo {author} {\bibfnamefont {M.~Y.}\ \bibnamefont
  {Kagan}},\ and\ \bibinfo {author} {\bibfnamefont {S.}~\bibnamefont
  {Stringari}},\ }\bibfield  {title} {\bibinfo {title} {Collective mode of
  homogeneous superfluid fermi gases in the bec-bcs crossover},\ }\href
  {https://doi.org/10.1103/PhysRevA.74.042717} {\bibfield  {journal} {\bibinfo
  {journal} {Phys. Rev. A}\ }\textbf {\bibinfo {volume} {74}},\ \bibinfo
  {pages} {042717} (\bibinfo {year} {2006})}\BibitemShut {NoStop}%
\bibitem [{\citenamefont {Zhao}\ \emph {et~al.}(2020)\citenamefont {Zhao},
  \citenamefont {Gao}, \citenamefont {Liang}, \citenamefont {Zou},\ and\
  \citenamefont {Yuan}}]{RPA}%
  \BibitemOpen
  \bibfield  {author} {\bibinfo {author} {\bibfnamefont {H.}~\bibnamefont
  {Zhao}}, \bibinfo {author} {\bibfnamefont {X.}~\bibnamefont {Gao}}, \bibinfo
  {author} {\bibfnamefont {W.}~\bibnamefont {Liang}}, \bibinfo {author}
  {\bibfnamefont {P.}~\bibnamefont {Zou}},\ and\ \bibinfo {author}
  {\bibfnamefont {F.}~\bibnamefont {Yuan}},\ }\bibfield  {title} {\bibinfo
  {title} {Dynamical structure factors of a two-dimensional fermi superfluid
  within random phase approximation},\ }\href
  {https://doi.org/10.1088/1367-2630/abab3d} {\bibfield  {journal} {\bibinfo
  {journal} {New Journal of Physics}\ }\textbf {\bibinfo {volume} {22}},\
  \bibinfo {pages} {093012} (\bibinfo {year} {2020})}\BibitemShut {NoStop}%
\bibitem [{\citenamefont {Minguzzi}\ \emph {et~al.}(2001)\citenamefont
  {Minguzzi}, \citenamefont {Ferrari},\ and\ \citenamefont
  {Castin}}]{RPA-weak}%
  \BibitemOpen
  \bibfield  {author} {\bibinfo {author} {\bibfnamefont {A.}~\bibnamefont
  {Minguzzi}}, \bibinfo {author} {\bibfnamefont {G.}~\bibnamefont {Ferrari}},\
  and\ \bibinfo {author} {\bibfnamefont {Y.}~\bibnamefont {Castin}},\
  }\bibfield  {title} {\bibinfo {title} {Dynamic structure factor of a
  superfluid fermi gas},\ }\href {https://doi.org/10.1007/s100530170036}
  {\bibfield  {journal} {\bibinfo  {journal} {The European Physical Journal D -
  Atomic, Molecular, Optical and Plasma Physics}\ }\textbf {\bibinfo {volume}
  {17}},\ \bibinfo {pages} {49} (\bibinfo {year} {2001})}\BibitemShut {NoStop}%
\bibitem [{\citenamefont {Bezvershenko}\ \emph {et~al.}(2021)\citenamefont
  {Bezvershenko}, \citenamefont {Halati}, \citenamefont {Sheikhan},
  \citenamefont {Kollath},\ and\ \citenamefont
  {Rosch}}]{BezvershenkoRosch2021}%
  \BibitemOpen
  \bibfield  {author} {\bibinfo {author} {\bibfnamefont {A.~V.}\ \bibnamefont
  {Bezvershenko}}, \bibinfo {author} {\bibfnamefont {C.-M.}\ \bibnamefont
  {Halati}}, \bibinfo {author} {\bibfnamefont {A.}~\bibnamefont {Sheikhan}},
  \bibinfo {author} {\bibfnamefont {C.}~\bibnamefont {Kollath}},\ and\ \bibinfo
  {author} {\bibfnamefont {A.}~\bibnamefont {Rosch}},\ }\bibfield  {title}
  {\bibinfo {title} {Dicke transition in open many-body systems determined by
  fluctuation effects},\ }\href
  {https://doi.org/10.1103/PhysRevLett.127.173606} {\bibfield  {journal}
  {\bibinfo  {journal} {Phys. Rev. Lett.}\ }\textbf {\bibinfo {volume} {127}},\
  \bibinfo {pages} {173606} (\bibinfo {year} {2021})}\BibitemShut {NoStop}%
\bibitem [{\citenamefont {J\"ager}\ \emph {et~al.}(2022)\citenamefont
  {J\"ager}, \citenamefont {Schmit}, \citenamefont {Morigi}, \citenamefont
  {Holland},\ and\ \citenamefont {Betzholz}}]{JagerBetzholz2022}%
  \BibitemOpen
  \bibfield  {author} {\bibinfo {author} {\bibfnamefont {S.~B.}\ \bibnamefont
  {J\"ager}}, \bibinfo {author} {\bibfnamefont {T.}~\bibnamefont {Schmit}},
  \bibinfo {author} {\bibfnamefont {G.}~\bibnamefont {Morigi}}, \bibinfo
  {author} {\bibfnamefont {M.~J.}\ \bibnamefont {Holland}},\ and\ \bibinfo
  {author} {\bibfnamefont {R.}~\bibnamefont {Betzholz}},\ }\bibfield  {title}
  {\bibinfo {title} {Lindblad master equations for quantum systems coupled to
  dissipative bosonic modes},\ }\href
  {https://doi.org/10.1103/PhysRevLett.129.063601} {\bibfield  {journal}
  {\bibinfo  {journal} {Phys. Rev. Lett.}\ }\textbf {\bibinfo {volume} {129}},\
  \bibinfo {pages} {063601} (\bibinfo {year} {2022})}\BibitemShut {NoStop}%
\bibitem [{\citenamefont {Halati}\ \emph
  {et~al.}(2020{\natexlab{a}})\citenamefont {Halati}, \citenamefont {Sheikhan},
  \citenamefont {Ritsch},\ and\ \citenamefont {Kollath}}]{HalatiKollath2020}%
  \BibitemOpen
  \bibfield  {author} {\bibinfo {author} {\bibfnamefont {C.-M.}\ \bibnamefont
  {Halati}}, \bibinfo {author} {\bibfnamefont {A.}~\bibnamefont {Sheikhan}},
  \bibinfo {author} {\bibfnamefont {H.}~\bibnamefont {Ritsch}},\ and\ \bibinfo
  {author} {\bibfnamefont {C.}~\bibnamefont {Kollath}},\ }\bibfield  {title}
  {\bibinfo {title} {Numerically exact treatment of many-body self-organization
  in a cavity},\ }\href {https://doi.org/10.1103/PhysRevLett.125.093604}
  {\bibfield  {journal} {\bibinfo  {journal} {Phys. Rev. Lett.}\ }\textbf
  {\bibinfo {volume} {125}},\ \bibinfo {pages} {093604} (\bibinfo {year}
  {2020}{\natexlab{a}})}\BibitemShut {NoStop}%
\bibitem [{\citenamefont {Halati}\ \emph
  {et~al.}(2020{\natexlab{b}})\citenamefont {Halati}, \citenamefont
  {Sheikhan},\ and\ \citenamefont {Kollath}}]{HalatiKollath2020b}%
  \BibitemOpen
  \bibfield  {author} {\bibinfo {author} {\bibfnamefont {C.-M.}\ \bibnamefont
  {Halati}}, \bibinfo {author} {\bibfnamefont {A.}~\bibnamefont {Sheikhan}},\
  and\ \bibinfo {author} {\bibfnamefont {C.}~\bibnamefont {Kollath}},\
  }\bibfield  {title} {\bibinfo {title} {Theoretical methods to treat a single
  dissipative bosonic mode coupled globally to an interacting many-body
  system},\ }\href {https://doi.org/10.1103/PhysRevResearch.2.043255}
  {\bibfield  {journal} {\bibinfo  {journal} {Phys. Rev. Research}\ }\textbf
  {\bibinfo {volume} {2}},\ \bibinfo {pages} {043255} (\bibinfo {year}
  {2020}{\natexlab{b}})}\BibitemShut {NoStop}%
\bibitem [{\citenamefont {Gardiner}\ and\ \citenamefont
  {Zoller}(2000)}]{GardinerZollerBook}%
  \BibitemOpen
  \bibfield  {author} {\bibinfo {author} {\bibfnamefont {C.}~\bibnamefont
  {Gardiner}}\ and\ \bibinfo {author} {\bibfnamefont {P.}~\bibnamefont
  {Zoller}},\ }\href@noop {} {\emph {\bibinfo {title} {Quantum Noise}}}\
  (\bibinfo  {publisher} {Spinger-Verlag},\ \bibinfo {year} {2000})\BibitemShut
  {NoStop}%
\bibitem [{\citenamefont {White}\ and\ \citenamefont
  {Feiguin}(2004)}]{WhiteFeiguin2004}%
  \BibitemOpen
  \bibfield  {author} {\bibinfo {author} {\bibfnamefont {S.~R.}\ \bibnamefont
  {White}}\ and\ \bibinfo {author} {\bibfnamefont {A.~E.}\ \bibnamefont
  {Feiguin}},\ }\bibfield  {title} {\bibinfo {title} {Real-time evolution using
  the density matrix renormalization group},\ }\href
  {https://doi.org/10.1103/PhysRevLett.93.076401} {\bibfield  {journal}
  {\bibinfo  {journal} {Phys. Rev. Lett.}\ }\textbf {\bibinfo {volume} {93}},\
  \bibinfo {pages} {076401} (\bibinfo {year} {2004})}\BibitemShut {NoStop}%
\bibitem [{\citenamefont {Daley}\ \emph {et~al.}(2004)\citenamefont {Daley},
  \citenamefont {Kollath}, \citenamefont {Schollwöck},\ and\ \citenamefont
  {Vidal}}]{DaleyVidal2004}%
  \BibitemOpen
  \bibfield  {author} {\bibinfo {author} {\bibfnamefont {A.~J.}\ \bibnamefont
  {Daley}}, \bibinfo {author} {\bibfnamefont {C.}~\bibnamefont {Kollath}},
  \bibinfo {author} {\bibfnamefont {U.}~\bibnamefont {Schollwöck}},\ and\
  \bibinfo {author} {\bibfnamefont {G.}~\bibnamefont {Vidal}},\ }\bibfield
  {title} {\bibinfo {title} {Time-dependent density-matrix
  renormalization-group using adaptive effective hilbert spaces},\ }\href
  {https://doi.org/10.1088/1742-5468/2004/04/P04005} {\bibfield  {journal}
  {\bibinfo  {journal} {Journal of Statistical Mechanics: Theory and
  Experiment}\ }\textbf {\bibinfo {volume} {2004}},\ \bibinfo {pages} {P04005}
  (\bibinfo {year} {2004})}\BibitemShut {NoStop}%
\bibitem [{\citenamefont {Schollw{\"o}ck}(2011)}]{Schollwoeck2011}%
  \BibitemOpen
  \bibfield  {author} {\bibinfo {author} {\bibfnamefont {U.}~\bibnamefont
  {Schollw{\"o}ck}},\ }\bibfield  {title} {\bibinfo {title} {The density-matrix
  renormalization group in the age of matrix product states},\ }\href
  {https://doi.org/http://dx.doi.org/10.1016/j.aop.2010.09.012} {\bibfield
  {journal} {\bibinfo  {journal} {Annals of Physics}\ }\textbf {\bibinfo
  {volume} {326}},\ \bibinfo {pages} {96 } (\bibinfo {year}
  {2011})}\BibitemShut {NoStop}%
\bibitem [{\citenamefont {White}(1992)}]{White1992}%
  \BibitemOpen
  \bibfield  {author} {\bibinfo {author} {\bibfnamefont {S.~R.}\ \bibnamefont
  {White}},\ }\bibfield  {title} {\bibinfo {title} {Density matrix formulation
  for quantum renormalization groups},\ }\href
  {https://doi.org/10.1103/PhysRevLett.69.2863} {\bibfield  {journal} {\bibinfo
   {journal} {Phys. Rev. Lett.}\ }\textbf {\bibinfo {volume} {69}},\ \bibinfo
  {pages} {2863} (\bibinfo {year} {1992})}\BibitemShut {NoStop}%
\bibitem [{\citenamefont {Fishman}\ \emph {et~al.}(2022)\citenamefont
  {Fishman}, \citenamefont {White},\ and\ \citenamefont
  {Stoudenmire}}]{FishmanStoudenmire2020}%
  \BibitemOpen
  \bibfield  {author} {\bibinfo {author} {\bibfnamefont {M.}~\bibnamefont
  {Fishman}}, \bibinfo {author} {\bibfnamefont {S.~R.}\ \bibnamefont {White}},\
  and\ \bibinfo {author} {\bibfnamefont {E.~M.}\ \bibnamefont {Stoudenmire}},\
  }\bibfield  {title} {\bibinfo {title} {{The ITensor Software Library for
  Tensor Network Calculations}},\ }\href
  {https://doi.org/10.21468/SciPostPhysCodeb.4} {\bibfield  {journal} {\bibinfo
   {journal} {SciPost Phys. Codebases}\ ,\ \bibinfo {pages} {4}} (\bibinfo
  {year} {2022})}\BibitemShut {NoStop}%
\bibitem [{\citenamefont {Shi}\ \emph {et~al.}(2018)\citenamefont {Shi},
  \citenamefont {Demler},\ and\ \citenamefont {{Ignacio Cirac}}}]{SHI2018245}%
  \BibitemOpen
  \bibfield  {author} {\bibinfo {author} {\bibfnamefont {T.}~\bibnamefont
  {Shi}}, \bibinfo {author} {\bibfnamefont {E.}~\bibnamefont {Demler}},\ and\
  \bibinfo {author} {\bibfnamefont {J.}~\bibnamefont {{Ignacio Cirac}}},\
  }\bibfield  {title} {\bibinfo {title} {Variational study of fermionic and
  bosonic systems with non-gaussian states: Theory and applications},\ }\href
  {https://doi.org/https://doi.org/10.1016/j.aop.2017.11.014} {\bibfield
  {journal} {\bibinfo  {journal} {Annals of Physics}\ }\textbf {\bibinfo
  {volume} {390}},\ \bibinfo {pages} {245} (\bibinfo {year}
  {2018})}\BibitemShut {NoStop}%
\end{thebibliography}%

\clearpage

\appendix

\section*{Materials and Methods}

\subsection*{Experimental procedure}

We produce a strongly interacting Fermi gas of ${}^6\mathrm{Li}$ comprising $5.68 \times 10^5$ atoms in the two lowest hyperfine states in the vicinity of the broad magnetic Feshbach resonance at $832\ \mathrm{G}$. 
We follow the method outlined in our previous work \cite{Roux:2020aa, roux:2021uf} with modifications in the last evaporation stages, leading to a deeply degenerate gas trapped in an elongated harmonic trap with transverse frequencies $\omega_{\mathrm{y, z}} = 430\ \mathrm{Hz}$ and a longitudinal frequency of $\omega_{\mathrm{x}} = 28\ \mathrm{Hz}$ along the cavity axis. 

As mentioned in the main text, the pump beam is operated in the atomic dispersive coupling regime, 
but very close to a $\mathrm{TEM}_{00}$ cavity mode. We monitor the power sent on the atoms during each shot on a photodiode. The pump potential depth is calibrated in atomic recoil units $E_\mathrm{R} = \hbar^2\mathbf{k}^2_{\mathrm{p}}/2m = h \times 73.67\ \mathrm{kHz}$ by first performing a Kapitza-Dirac diffraction experiment on a molecular BEC at $690\ \mathrm{G}$ using a standing-wave pump beam with $\pi$-polarization, obtaining the pump lattice depth $V_{0, \mathrm{KD}}$ \cite{GadwayAOKDDPBRNR2009}. We subsequently cancel the pump lattice by introducing linear, orthogonal polarizations for the pump beam and its back-reflection as shown in Fig.\ref{fig:exp_setup}A, which we verify through the absence of diffraction peaks in the Kapitza-Dirac experiment. The geometry and properties of the pump are otherwise the same as in \cite{HelsonDWOIAUFGWPMI2023}. The lattice-free pump potential depth is given by $V_0 = V_{0, \mathrm{KD}}/2$. 

The long-range interaction strength $D_0 \propto V_0/\tilde{\Delta}_{\mathrm{c}}$ can be tuned by varying the pump detuning from the dispersively-shifted cavity $\tilde{\Delta}_{\mathrm{c}}$ or the pump potential depth $V_0$. The pump-cavity detuning $\tilde{\Delta}_{\mathrm{c}} = \Delta_{\mathrm{c}} - \delta_{\mathrm{c}}$ is chosen to largely exceed the cavity dispersive shift of $\delta_{\mathrm{c}} = - 2 \pi \times 1.5\ \mathrm{MHz}$ and the cavity linewidth of $\kappa = 2 \pi \times 77\ \mathrm{kHz}$. For sudden quenches, we switch on $V_0$ from zero to a finite value on a characteristic timescale of $1\ \mu\mathrm{s}$, which is faster than the dissipative photon dynamics on the timescale $1/\kappa \approx 2\ \mu\mathrm{s}$ and any atomic motion in the uniform state at $\hbar/E_{\mathrm{F}} \approx 7.7\ \mu\mathrm{s}$, but slower than the coherent photon dynamics at $2\pi/\tilde{\Delta}_{\mathrm{c}}$.

After a quench, the photon flux leaking from the cavity is detected using a single-photon counter with an efficiency
of approximately 3\% \cite{HelsonOROASIFG2022}.

\subsection*{Data analysis}

\subsubsection*{Analysis of sudden quenches}
\begin{figure}[t!]
 \includegraphics{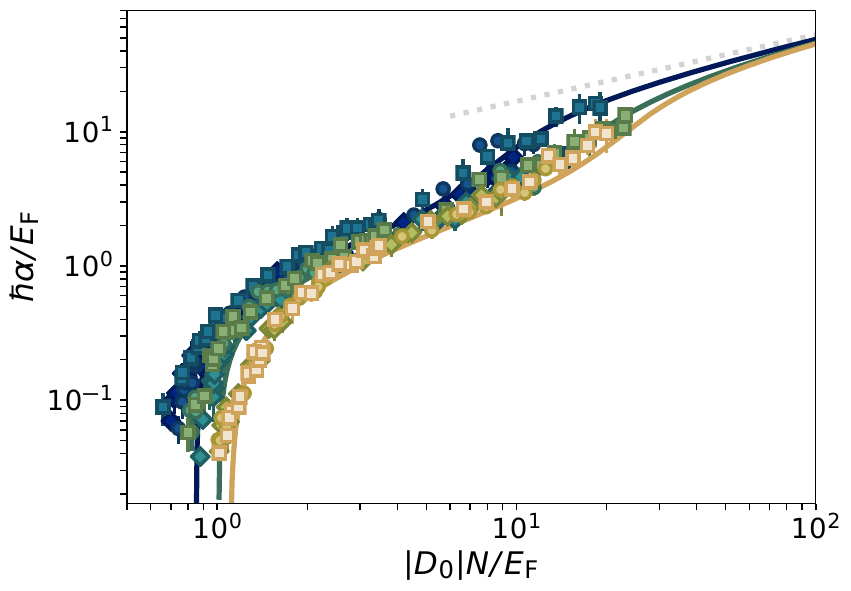}
\caption{\textbf{Growth rates in the BEC-BCS Crossover.}
Growth rates $\alpha$ in units of the Fermi energy $E_{\mathrm{F}}$ in the BEC-BCS Crossover as a function of $D_0$ for different pump-cavity detunings $\tilde{\Delta}_{\mathrm{c}}$: $-3.4\ \mathrm{MHz}$ ($ \textcolor{grey1}{\meddiamond}$), $-2.5\ \mathrm{MHz}$ ($ \textcolor{grey1}{\medcircle}$), $-1.4\ \mathrm{MHz}$ ($ \textcolor{grey1}{\medsquare}$) and $1/k_{\mathrm{F}}a$: $1.1$ ($\textcolor{batlowS1}{\meddiamond} \textcolor{batlowS2}{\medcircle} \textcolor{batlowS3}{\medsquare}$), $0$ ($\textcolor{batlowS4}{\meddiamond} \textcolor{batlowS5}{\medcircle} \textcolor{batlowS6}{\medsquare}$), $-0.7$ ($ \textcolor{batlowS7}{\meddiamond} \textcolor{batlowS8}{\medcircle} \textcolor{batlowS9}{\medsquare}$). The results of an instability analysis for BEC (\textcolor{batlowS3}{\sampleline{}}), UFG (\textcolor{batlowS6}{\sampleline{}}) and BCS (\textcolor{batlowS9}{\sampleline{}}) show very good agreement with the experimental data. For $D_0 \rightarrow \infty$, the growth rates tend toward the universal value given by the f-sum rule (\textcolor{grey1}{\sampleline{dotted}}).
\label{sup:growth rate}}
\end{figure}

We extract the system response to sudden quenches by performing an exponential fit on the histograms of photon counts in the early time for each experimental realization. We apply a moving average before fitting to mitigate photon shot noise. The fit function is given by:
\begin{equation}
\bar{n}_{\mathrm{det}}(t) = \bar{n}_0 e^{\alpha t},
\end{equation}
with a seed flux value $\bar{n}_0$ and a growth rate $\alpha$. The fit range is up to either $60\%$ of the maximal value or a maximal count rate of $20\ \mathrm{MHz}$, depending on which value is reached first. At count rates larger than $20\ \mathrm{MHz}$, the photon flux traces are significantly distorted due to dead time effects of the single-photon counter. The fit is performed as a linear fit on a log-scale, therefore equally weighting low and high count rates in the least-square minimization. In Fig.~\ref{sup:growth rate}, the full dataset of normalized growth rates $\hbar \alpha/E_{\mathrm{F}}$ as a function of $D_0$ for three different detunings $\Tilde{\Delta}_\mathrm{c}$ and short-range interaction strengths $1/k_\mathrm{F}a$ is shown. The data shows very good quantitative agreement with the respective theory curves obtained from the instability analysis using a trap-averaged random-phase approximation density response function. At very large $D_0$ the growth rate tends toward a universal value obtained from the f-sum rule as described in the main text, which is independent of the microscopic details of the atomic gas.

\subsubsection*{Analysis of dynamic ramps} 

We extract the apparent ordering threshold $D_{0,\textrm{thr}}$, at which the photon rate $\bar{n}_{\mathrm{det}}$ reaches $1\,$MHz, by fitting the onset of the acquired photon flux traces with a linear function:
\begin{equation}
    \bar{n}_{\mathrm{det}}(t) = \theta (t-t_0)  B(t-t_0),
\end{equation}
where $\theta(t)$ is the Heaviside function, $t_0$ and $B$ fit parameters. From this fit we obtain the time $t_\mathrm{thr}$ at which the photon flux exceeds $1\,$MHz, which is subsequently converted to a long-range interaction strength, $D_{0,\mathrm{thr}}$ using the linear ramp shape.
For this we perform a linear fit to the pump power, which is monitored for each realization on a photodiode, and use it to extract its power at $t_\mathrm{thr}$, which we then convert into $D_0$. The same linear fit on the photodiode signal provides also a measurement of the ramp speed $\dot{\epsilon}$.

To extract the scaling exponent $\gamma$, we perform a power-law fit of $\epsilon_{\mathrm{thr}}$ over a limited range of ramp speed $\dot{\epsilon}$ (see Fig.~\ref{fig:ramps} D of the main text). In particular, we constrain the fit region up to $\dot{\epsilon} = 200\ \mathrm{ms}^{-1}$, which is the largest ramp speed at $1/k_\mathrm{F}a = -0.7$. On the other hand, $\epsilon_{\mathrm{thr}}$ can be artificially increased by atom losses happening before reaching the threshold in the low speed limit (this is strongest for the data at $1/k_\mathrm{F}a = 1.1$ in Fig.~\ref{fig:ramps} C), because of the long duration of the pump power ramp. We exclude the loss-dominated region at low speed by performing a two-step fit: we first fit $\epsilon_{\mathrm{thr}}$ over the full range of $\dot{\epsilon}$, then exclude all the points up to the $\dot{\epsilon}$ value where the residual of the first fit are 25\%. The power-law fit performed over such a restricted range of $\dot{\epsilon}$ provides the best value of $\gamma$ and its corresponding fit error is typically on the order of $5 \,\%$.

The method we used to obtain the optimal value of $\gamma$ is subject to two sources of systematic error. First, both $\epsilon_{\mathrm{thr}}$ and $\dot{\epsilon}$ are calculated from the measured value of $D_{0\mathrm{C}}$, which we extract as described above. To account for the systematic on $\gamma$ reflected by the error on $D_{0\mathrm{C}}$, we repeat the power-law fit on $\epsilon_{\mathrm{thr}} (\dot{\epsilon})$ several times, employing for each fit a different value of the critical long-range interaction strength in the confidence range of the measured $D_{0\mathrm{C}}$. Calculating the standard deviation over the values of $\gamma$ obtained from the different fits, we estimate the systematic error on the power-law exponent due to the finite precision on measuring $D_{0\mathrm{C}}$ to be on the order of $\sim 1\ \%$ for all the datasets. 
Second, a source of systematic error on $\gamma$ arises from our selection of the range of $\dot{\epsilon}$ over which the fit is conducted. We estimate this contribution by releasing the fit range constraints and computing the standard deviation over the different values of $\gamma$ obtained in the extended range of $\dot{\epsilon}$. In particular, to estimate the systematic, we let the fit range to vary up to $1000\,$ms$^{-1}$, corresponding to the absolute largest ramp speed, 
and down to the value of $\dot{\epsilon}$ where the residual of the first fit are $10\ \%$.
We estimate the systematic error on $\gamma$ due to choice of the fit range to be of the order of $\sim 7\ \%$ for all the datasets. The final error on $\gamma$ is then calculated as the quadratic sum of the different error contributions: the fitting error on the best fit and the two systematic errors. The value of $\gamma = 0.731(16)$  reported in the main text corresponds to the weighted average, and weighted standard deviation, over the different critical exponents obtained for different $\tilde{\Delta_c}$ and short-range interaction strength.

\subsection*{Effective Hamiltonian}

In this section, we derive the effective Hamiltonian describing the transversely pumped atom-cavity system. The Fermi gas inside the optical resonator is transversely pumped by a pump laser with a geometry as shown in Fig.~\ref{fig:exp_setup}. The pump beam and its back-reflection are both in a coherent state with equal amplitude $\lambda_{\mathrm{p}}$. Their linear and orthogonal polarizations are described by the unit vectors $\mathbf{e}_1 = 1/\sqrt{2}(\mathbf{e}_{\mathrm{z}} - \mathbf{e}_{\mathrm{p}})$ and $\mathbf{e}_2 = 1/\sqrt{2}(\mathbf{e}_{\mathrm{z}} + \mathbf{e}_{\mathrm{p}})$ with the polarization vector in the pump-cavity plane $\mathbf{e}_{\mathrm{p}} = \cos(\phi) \mathbf{e}_{\mathrm{y}} + \sin(\phi) \mathbf{e}_{\mathrm{x}}$ and the angle of the pump to the cavity axis $\phi = \pi/10$. The atoms couple to two degenerate, standing-wave modes of the cavity described by the annihilation operators $\hat{a}_{{\mathrm{y/z}}}$ with respective polarizations along y and z. The combined pump-cavity field is given by:

\begin{align}
{\phi}(\mathbf{\mathbf{r}})
&= \cos(\mathbf{k}_\mathrm{c} \mathbf{r})(\hat{a}_\mathrm{y} \mathbf{e}_\mathrm{y} + \hat{a}_\mathrm{z} \mathbf{e}_\mathrm{z}) \\
&+ \lambda_{\mathrm{p}}(e^{i\mathbf{k}_\mathrm{p}\mathbf{r}} \mathbf{e}_\mathrm{1} + e^{-i\mathbf{k}_\mathrm{p}\mathbf{r}} \mathbf{e}_\mathrm{2}). \nonumber
\end{align}

The absence of vector polarizability in our experimental regime leads to the usual dispersive light-matter coupling of atomic density of both hyperfine states $\hat{n}(\mathbf{r}) = \sum_{\sigma=\uparrow\downarrow}\hat{n}_{\sigma}(\mathbf{r})$ to an effective pump-cavity potential \cite{mivehvar:2021aa} given by:

\begin{equation}
    H_\mathrm{int} = \int d\mathbf{r} \hat{n}(\mathbf{r}) \hat{V}_{\mathrm{eff}}(\mathbf{r}),
\end{equation}

where:

\begin{align}
    \hat{V}_{\mathrm{eff}}
    &= V_0 + U_0(\hat{a}^{\dagger}_\mathrm{y}\hat{a}_\mathrm{y}+\hat{a}^{\dagger}_\mathrm{z}\hat{a}_\mathrm{z})\cos^2(\mathbf{k}_\mathrm{c}\mathbf{r}) \\
    &+ \eta_0 [i\cos(\phi) \cos(\mathbf{k}_\mathrm{c}\mathbf{r})\sin(\mathbf{k}_\mathrm{p}\mathbf{r}) (\hat{a}_\mathrm{y}-\hat{a}^{\dagger}_\mathrm{y}) \nonumber \\
    &+ \cos(\mathbf{k}_\mathrm{c}\mathbf{r})\cos(\mathbf{k}_\mathrm{p}\mathbf{r}) (\hat{a}_\mathrm{z}+\hat{a}^{\dagger}_\mathrm{z})], \nonumber
\end{align}

with lattice-free pump potential of depth $V_0$, the intra-cavity lattice depth per photon $U_0$ and $\eta_0 = \sqrt{U_0 V_0}$.
The full Hamiltonian can then be recast in the form:

\begin{align}
\hat{H} 
&= \hat{H}_{\mathrm{at}} - \tilde{\Delta}_\mathrm{c}(\hat{a}^{\dagger}_\mathrm{y}\hat{a}_\mathrm{y}+\hat{a}^{\dagger}_\mathrm{z}\hat{a}_\mathrm{z}) \\
&+\eta_0 [\cos(\phi) \hat{\Theta}_y (\hat{a}_y-\hat{a}^{\dagger}_y) + \hat{\Theta}_z (\hat{a}_z+\hat{a}^{\dagger}_z)], \nonumber 
\end{align}

where $\tilde{\Delta}_\mathrm{c} = \Delta_\mathrm{c} - \delta_\mathrm{c} = \Delta_\mathrm{c} - U_0 \int d\mathbf{r} n(\mathbf{r}) \cos^2(\mathbf{k}_\mathrm{c}\mathbf{r})$ is the detuning of the pump from the dispersively-shifted cavity resonance and $\hat{H}_{\mathrm{at}}$ is the Hamiltonian of an interacting, harmonically trapped two-component Fermi gas. The potential offset created by the pump has been dropped.
Furthermore, we introduce the density-wave operators for polarizations $y$ and $z$, describing the modulation of the atomic density at wave vectors $\mathbf{k}_{\pm} = \mathbf{k}_\mathrm{p} \pm \mathbf{k}_\mathrm{c}$ :
\begin{align}
    \hat{\Theta}_\mathrm{y}
    &= \int d\mathbf{r}\cos(\mathbf{k}_\mathrm{c}\mathbf{r})\sin(\mathbf{k}_\mathrm{p}\mathbf{r}) \hat{n}(\mathbf{r}) \\
    &= \frac{1}{4}(\hat{n}_{\mathrm{k}_-}-\hat{n}_{-\mathrm{k}_-}-\hat{n}_{\mathrm{k}_+}+\hat{n}_{-\mathrm{k}_+}) \nonumber \\
    \hat{\Theta}_\mathrm{z} 
    &= \int d\mathbf{r} \cos(\mathbf{k}_\mathrm{c}\mathbf{r})\cos(\mathbf{k}_\mathrm{p}\mathbf{r})\hat{n}(\mathbf{r}) \nonumber \\
    &= \frac{1}{4}(\hat{n}_{\mathrm{k}_-}+\hat{n}_{-\mathrm{k}_-}+\hat{n}_{\mathrm{k}_+}+\hat{n}_{-\mathrm{k}_+}), \nonumber
\end{align}
with $\hat{n}_\mathrm{q} = \int d\mathbf{r} \hat{n}(\mathbf{r})e^{i\mathbf{q}\mathbf{r}}$ being the Fourier transform of the fermionic density. Since $\Tilde{\Delta}_\mathrm{c}$ is the largest energy scale, ensuring that the coherent cavity dynamics are much faster than any other processes, the cavity photon field can be integrated out. This results in an effective long-range interaction Hamiltonian:
\begin{equation}
    H = H_{\mathrm{at}} + \int d \mathbf{r} d \mathbf{r}' D(\mathbf{r},\mathbf{r}') \hat{n}(\mathbf{r}) \hat{n}(\mathbf{r}'),
\end{equation}
where the effective long-range interaction $D(\mathbf{r},\mathbf{r}')$ is given in Eq.~\eqref{eqn:Interaction}:
\begin{multline}
        D(\mathbf{r},\mathbf{r}')=D_0\left[ \cos \left(\mathbf{k}_+ \cdot (\mathbf{r} -\mathbf{r}' ) \right) + \cos \left(\mathbf{k}_- \cdot (\mathbf{r} -\mathbf{r}' ) \right) +\right.\\ \left. \cos \left(\mathbf{k}_+\mathbf{r} + \mathbf{k}_-\mathbf{r}' \right) + \cos \left(\mathbf{k}_-\mathbf{r} + \mathbf{k}_+\mathbf{r}' \right) \right],
\end{multline}
with the long-range interaction strength $D_0 = \eta_0^2/\Tilde{\Delta}_\mathrm{c} = U_0V_0/\Tilde{\Delta}_\mathrm{c}$.
We note that the two cavity modes couple to linearly independent combinations of DW operators $\hat{\Theta}_y$ and $\hat{\Theta}_z$, hence within linearized equations of motion the two fields remain uncoupled. The main difference between the two terms is that the term $\hat{\Theta}_z$ couples to $\eta_0$ and the term $\hat{\Theta}_x$ couples to $\eta_0 \cos\phi $. In the following, we analyze the coupling of the fermionic atoms to only one cavity mode, as including the other is a straightforward extension. 

\subsection*{Instability analysis and equations of motion}

In this section, we provide details on the general derivation of the growth rate equation (Eq.~\eqref{eq:theory}). For a single cavity mode, the Hamiltonian can be written as:
        \begin{equation}
        H =H_{\mathrm{at}} + \frac{\Tilde{\Delta}_\mathrm{c}}{2} (\hat{x}^2 + \hat{p}^2) +  \sqrt{2} \eta(t) x \hat{\Theta},
        \label{eq:hamiltonian}
    \end{equation}
with $x,p$ being the canonical quadrature operators $ \hat{x}= \frac{1}{\sqrt{2}}(\hat{a}^\dag + \hat{a})$ and $\hat{p} = \frac{i}{\sqrt{2}}(\hat{a}^\dag - \hat{a})$. Here, $\hat{\Theta} = \hat{\Theta}_\mathrm{z}$ and long-range coupling strength $\eta(t) = \eta_0$ for $t > 0$ and $\eta(t) = 0$ before the quench. To compute the growth rate, we begin by using the linearized Heisenberg equations of motion, which for the photon field are
\begin{equation}
    \dot{\hat{x}} = -\Tilde{\Delta}_\mathrm{c} \hat{p}, ~~~ \dot{\hat{p}} = \Tilde{\Delta}_\mathrm{c} \hat{x} - \sqrt{2}\Tilde{\Delta}_\mathrm{c} \eta_0 \hat{\Theta}.  
\label{eq:heisenberg}
\end{equation}

Computing the equation of motion for the DW operator is complex, as its exact form depends on the microscopic details. However, we can utilize linear response theory to express the expectation value of the density operator: 
\begin{equation}
    n_\mathrm{q}(t) = \frac{\sqrt{2}}{4}\eta_0 N \int_{-\infty}^\infty dt' \chi(q,t-t') x(t'),
\label{eq:LinRespTime}
\end{equation}
where $N$ is the total number of fermions in the cavity and $\chi(q,t)$ is the standard (intensive) density-density response function of the interacting fermions in absence of the cavity \cite{Dupuis}. Combining Eqs.~\eqref{eq:heisenberg}~-~\eqref{eq:LinRespTime}, we obtain a second order integro-differential equation for the expectation value of $x$:
\begin{equation}
    \Ddot{x} + \Tilde{\Delta}_c^2 x -  \frac{1}{8}\eta_0^2 N \Tilde{\Delta}_c \int_{-\infty}^\infty dt' \sum_{q=\pm \mathrm{k}_{\pm}} \chi(q,t-t') x(t') = 0.
\end{equation}

As we are looking for quench instabilities, we consider an exponentially growing solution of the form $x(t) = x(0)e^{\alpha t}$ as an \textit{ansatz}. This allows us to obtain a non-linear, self-consistent equation for the instability growth rate $\alpha$:
\begin{equation}
    \alpha^2 + \Tilde{\Delta}_c^2  - \frac{1}{8}\eta_0^2 N \Tilde{\Delta}_c \sum_{q = \pm \mathrm{k}_\pm} \chi(q, \omega = i\alpha) = 0 
\end{equation}
Since the photon detuning is the largest energy scale, we neglect the $\alpha^2$ term in the above equation, leaving $D_0$ as the only relevant experimental parameter. The final instability growth rate equation is:

\begin{equation}
   1 = \frac{1}{8} D_0 N \sum_{q= \pm \mathrm{k}_{\pm}} \chi(q, \omega = i\alpha).
\end{equation}

By neglecting the $\alpha^2$-term, we simplify the dynamics, implying that the behavior of the cavity photons is not relevant to the instability. Instead, photons simply mediate the long-range interactions between fermions, with an amplitude determined by $D_0$. The above expression is equivalent to Eq.~\eqref{eq:theory} by re-expressing $\chi(q,\omega = i\alpha)$ as an integral over real-frequencies via the spectral representation \cite{Dupuis}.

\subsubsection*{RPA density response in the BEC-BCS Crossover}

In this section, we outline the derivation of the interacting fermion response function within random-phase approximation (RPA), which captures the collective sound mode of the system, as well as the usual particle-hole continuum, which is gapped \cite{Response-sound}. We follow the derivation of Refs.~\cite{RPA,he:2016aa,RPA-weak}, where the approach is based on computing equations of motion of individual operator and then including the self-consistent field in a canonical way. The atomic Hamiltonian in momentum space reads:
\begin{equation}
    H_{\mathrm{at}} = \sum_k (\xi_k-\mu) \hat{c}^\dag_k \hat{c}_k + U \sum_{k,k',q,\sigma,\sigma'} \hat{c}^\dag_{k+q, \sigma} \hat{c}^\dag_{k'-q,\sigma'} \hat{c}_{k',\sigma'} \hat{c}_{k,\sigma},
\end{equation}
where $\xi_k$ is the single fermion dispersion and $U$ is an effective interaction strength, which satisfies the usual T-matrix/scattering length equation:

\begin{equation}
    \frac{m}{4\pi a_s} = \frac{1}{U} + \frac{1}{V} \sum_\mathrm{k}^{k_\mathrm{c}} \frac{m}{k^2}.
    \label{eqn:scattering_length}
\end{equation}

We proceed by performing a mean-field decoupling and derive the BCS Hamiltonian:

\begin{equation}
    H^0 = \sum_{k,\sigma} (\xi_k - \mu) \hat{c}^\dag_{k,\sigma} \hat{c}_{k,\sigma} + \Delta \sum_k \hat{c}^\dag_{k,\uparrow} \hat{c}^\dag_{k,\downarrow} + \text{H.c.} 
\end{equation}

Using the Hamiltonian $H^0$, we can compute the individual response functions $\chi^0$ due to external driving. To include the RPA response, we employ mean-field decoupling of the interacting Hamiltonian, but also keep any term proportional to the expectation value of the operator we are driving for example $\langle{\hat{n}_q}\rangle$. Overall, this gives rise to the RPA response function:

\begin{equation}
    \chi^{\mathrm{RPA}} = \frac{\chi^0}{1 - U \chi^0 (\sigma_\mathrm{x} \oplus \sigma_\mathrm{x})},
\end{equation}

where $\sigma_\mathrm{x}$ is a Pauli matrix. The above matrix equation should be understood as if acting on the vector $(\delta \hat{n}_{q\uparrow}, \delta \hat{n}_{q\downarrow}, \delta \hat{\Delta}, \delta \hat{\Delta}^\dag)$, with $\hat{\Delta}^\dag_q = \sum_k \hat{c}^\dag_{k+q\uparrow}\hat{c}^\dag_{-k\downarrow}$. By evaluating the entries and performing the matrix product, we obtain the expression for the density-density response of a homogeneous system.
The above expression can now be generalized to the trap-averaged case by replacing it with \cite{RPA-weak}:

\begin{equation}
    \chi^{\mathrm{TA}}(q,\omega) = \int d\mathbf{r} n(\mathbf{r}) \chi^{\mathrm{RPA}}(q,\omega, n(\mathbf{r}), \Delta(\mathbf{r})), 
\end{equation}

where $n(r)$ is the local density and $\Delta(r)$ is the local superfluid gap. $n(\mathbf{r})$ satisfies $\int d\mathbf{r} n(\mathbf{r}) = N$, with $N$ being the total atom number. We use this result in Eq.~\eqref{eq:theory} to obtain the growth rate $\alpha$.

\subsection*{Numerical simulations for simplified systems}

\subsubsection*{tMPS calculations in 1D}

\begin{figure}[ht!]
 \includegraphics[]{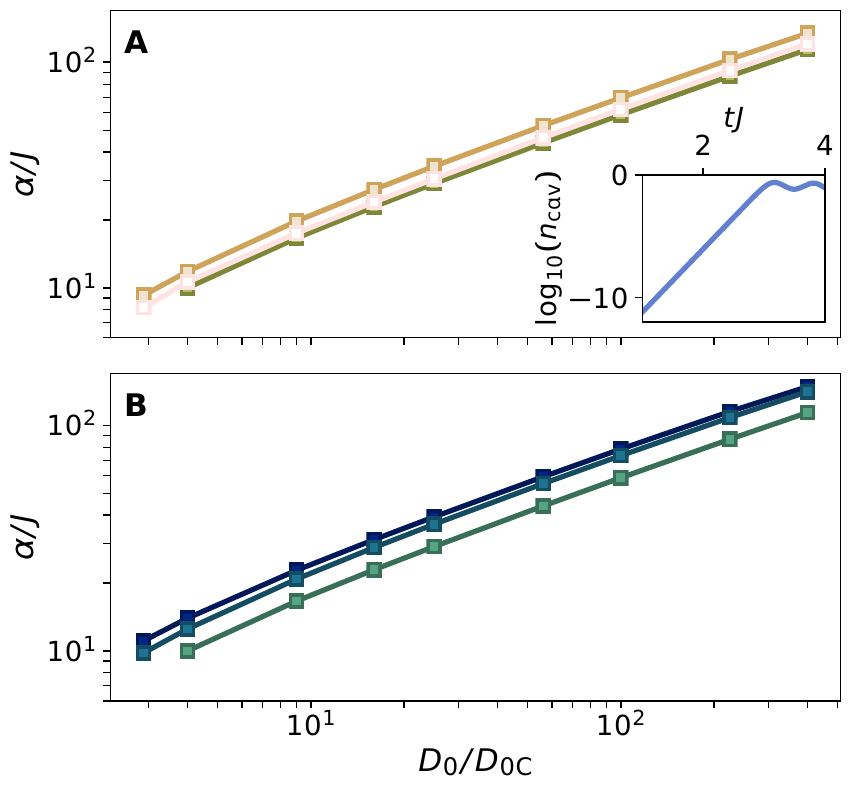}
\caption{\textbf{One-dimensional cavity-atoms simulations.} \textbf{(A)} Growth rate $\alpha$ in units of the hopping amplitude $J$ as a function of $D_0/D_{0\mathrm{C}}$, obtained from the numerical simulations of Eqs.~(\eqref{eq:Hamiltonian_1D})~-~(\eqref{eq:mf_photon_field}) for fermionic atoms, Eq.~(\eqref{eq:Hamiltonian_fermions}), with on-site interaction strengths $U/J$:  0 (\textcolor{batlowS7}{\sampleline{}}), 4 (\textcolor{batlowS9}{\sampleline{}}), 16 (\textcolor{batlowS12}{\sampleline{}}). The inset shows the exponential growth of the intracavity photon number $n_{\mathrm{cav}}$ for a quench to $D_0/D_{0\mathrm{C}}=4$ and $U/J = 4$ as a function of time $t$. \textbf{(B)} Growth rates for bosonic atoms, Eq.~(\eqref{eq:Hamiltonian_bosons}), with $U/J$: 0.25 (\textcolor{batlowS1}{\sampleline{}}), 4 (\textcolor{batlowS3}{\sampleline{}}), $\infty$ (\textcolor{batlowS5}{\sampleline{}}). 
The parameters used for panels A and B are $\hbar\Tilde{\Delta}_c/J=200$, $\hbar\kappa/J=3.75$, considering a similar separation of scales as in the experiment. 
The simulations were performed for $L=48$ sites and an atomic filling $N/L=0.375$. The filling was chosen to not be commensurate with the potential induced by the cavity.}
\label{fig:sim_tmps_classical_field}
\end{figure}

In this section, we show that the picture of the ordering dynamics has applicability beyond the three dimensional Fermi gas described in the main text. In this regard, we study coupled cavity-atoms systems, in which the atoms are confined to a one-dimensional lattice. 
In agreement with the experimental observations, we obtain an exponential ordering dynamics at early times, whose growth rate is mainly influenced by the strength of the long-range interactions. 

In the following we perform numerical simulations of one-dimensional interacting systems coupled to the field of the optical cavity.
We recover the short time-dynamics of interest by employing a mean-field description of the cavity field. 
This approach neglects the fluctuations in the atoms-cavity coupling \cite{mivehvar:2021aa, BezvershenkoRosch2021, JagerBetzholz2022}, however it allows us to access larger sizes for the atomic system than in the simulations which treat the cavity field exactly \cite{HalatiKollath2020,HalatiKollath2020b}.
Thus, we consider the following Hamiltonian for the coupled atoms-cavity system \cite{Ritsch:2013aa, mivehvar:2021aa}:

\begin{align} 
\label{eq:Hamiltonian_1D}
&H=H_{\text{atoms}}+H_{\text{ac}} \\
&H_{\text{ac}}=  -\hbar\eta \left( \lambda_{\mathrm{c}}^*+ \lambda_{\mathrm{c}}\right) \sum_j\cos(k_c j) \hat{n}_j \nonumber,
\end{align}

where $\hat{n}_j$ is the atomic density operator and $\lambda_{\mathrm{c}}$ is a complex number describing the cavity field. The equation of motion of the cavity field is given by:

\begin{align} 
\label{eq:mf_photon_field}
&\frac{\partial}{\partial t} \lambda_{\mathrm{c}} = i \eta \sum_j\cos(k_\mathrm{c} j) \left\langle \hat{n}_j\right\rangle-\left(i\Tilde{\Delta}_\mathrm{c}+\kappa\right)\lambda_{\mathrm{c}}.
\end{align}

For the results presented in this section, we consider the case in which the cavity is transversely pumped, with $\Tilde{\Delta}_\mathrm{c}$ the detuning of the cavity with respect to the pump beam and the strength of the cavity atoms effective coupling is given by $\eta$. 
The photon losses are controlled by the dissipation strength $\kappa$. As we are interested in the short-time dynamics, we did not include a stochastic Langevin noise term in Eq.~(\eqref{eq:mf_photon_field}) \cite{GardinerZollerBook}, which would be relevant at later times.
We choose the commensurability between the cavity wave vector and the one-dimensional lattice such that $k_c=\pi$, i.e.~the cavity couples to the atomic odd-even density imbalance, $\sum_j (-1)^j \left\langle \hat{n}_j\right\rangle$.

To further emphasize the generality of our conclusions, for the atomic Hamiltonian we consider interacting models for both spinful fermionic atoms:

\begin{align} 
\label{eq:Hamiltonian_fermions}
&H_\mathrm{F}=H_{\mathrm{int}}+H_{\mathrm{kin}} \\
&H_{\mathrm{int}}=U \sum_{j} \hat{n}_{j,\uparrow}\hat{n}_{j,\downarrow},\nonumber\\
&H_{\mathrm{kin}}=-J \sum_{j,\sigma} (\hat{c}_{j,\sigma}^\dagger \hat{c}_{j+1,\sigma} + \hat{c}_{j+1,\sigma}^\dagger \hat{c}_{j,\sigma}), \nonumber
\end{align}

and bosonic atoms:

\begin{align} 
\label{eq:Hamiltonian_bosons}
&H_\mathrm{B}=H_{\mathrm{int}}+H_{\mathrm{kin}} \\
&H_{\mathrm{int}}=\frac{U}{2} \sum_{j=1}^L \hat{n}_{j}(\hat{n}_{j}-1),\nonumber\\
&H_{\mathrm{kin}}=-J \sum_{j=1}^{L-1} (\hat{b}_{j}^\dagger \hat{b}_{j+1} + \hat{b}_{j+1}^\dagger \hat{b}_{j}). \nonumber
\end{align}

In both cases $J$ represents the tunneling amplitude and $U$ the on-site repulsive interaction strength.
These models in the presence of the cavity induced global-range interactions exhibit a phase transition to a density wave ordered state in their steady states as the atoms-cavity coupling is increased \cite{Ritsch:2013aa, mivehvar:2021aa}. 

While the cavity is treated at a mean-field level, we simulate the atomic dynamics numerically exactly. We employ a time-dependent matrix product state (tMPS) algorithm \cite{WhiteFeiguin2004, DaleyVidal2004, Schollwoeck2011} for the Hamiltonian in Eq.~(\ref{eq:Hamiltonian_1D}), while numerically integrating in parallel Eq.~(\ref{eq:mf_photon_field}).
The initial state of the evolution corresponds to a state without any photons, $\lambda_\mathrm{c}(t=0)=0$ and the atoms in the ground state of the model. The ground state is computed using the density matrix renormalization group (DMRG) algorithm in the matrix product state (MPS) representation \cite{White1992, Schollwoeck2011}. The implementations make use of the ITensor Library \cite{FishmanStoudenmire2020}. 

We focus on the early times of the coupled atom-cavity dynamics, where we observe an exponential increase in the photon number, $n_\mathrm{cav}=|\lambda_\mathrm{c}|^2$. 
We obtain the exponential dynamics for all parameters considered and regardless of the statistics of the atomic model.
We extract the growth rate of the exponential for different values of the atomic interactions and coupling $\eta$, presented in Fig.~\ref{fig:sim_tmps_classical_field}, for parameters with similar scale separation as in the experiment.
The critical value $\eta_\mathrm{c}$ has been determined by analyzing the self-organization phase transition, which occurs in the ground state of the Hamiltonian Eq.~(\ref{eq:Hamiltonian_1D}), computed self-consistently with the mean-field relations derived from the steady state of Eq.~(\ref{eq:mf_photon_field}).
We observe that the main influence on the growth rate stems from strength of the coupling to the cavity field $\eta$.
Interestingly, we obtain very similar growth rates for both bosonic and fermionic atoms, for all values of the short-range interactions considered, ranging from non-interacting to strongly repulsive.
This further suggests that the exponential growth of the density wave order is mostly controlled by the strength of the long-range interactions, while being only weakly influenced by the microscopic details of the atoms.

\subsubsection*{Gaussian state calculations in 2D}

\begin{figure}
\includegraphics[width=\columnwidth]{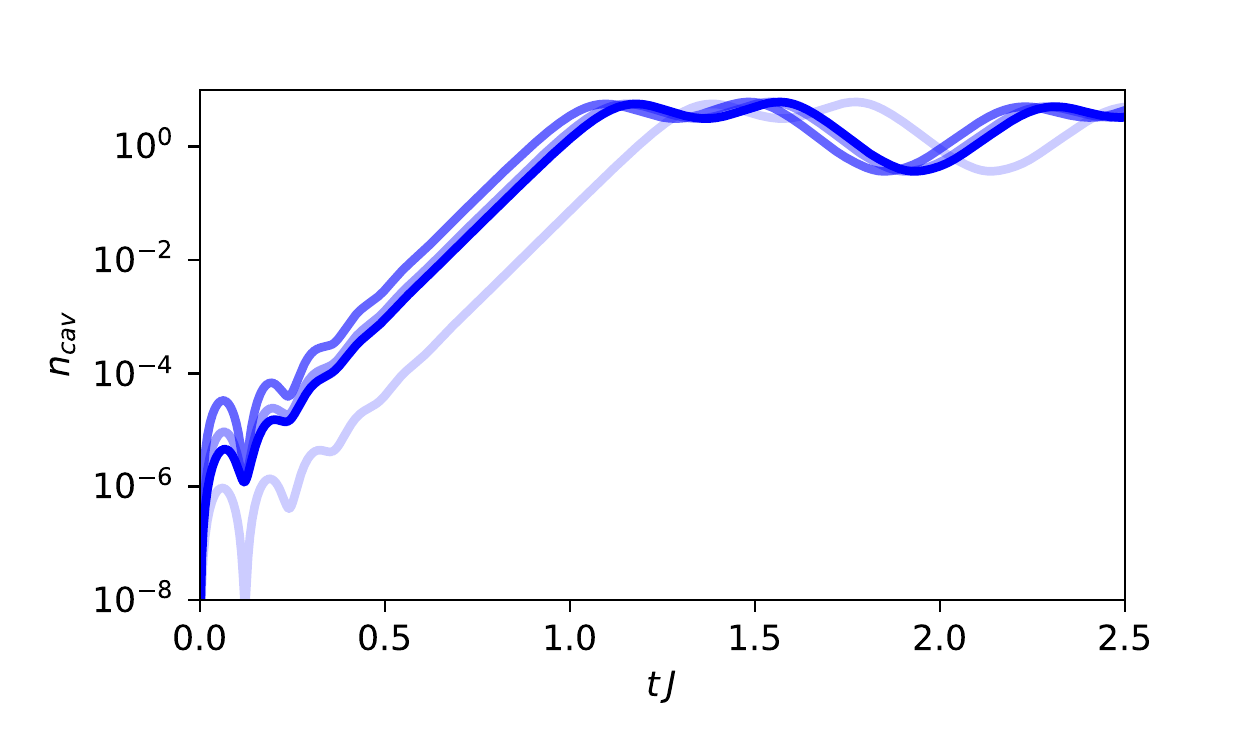}
\caption{\textbf{Two-dimensional Gaussian state simulation.} Intracavity photon number $n_{cav}$ vs. time $t$ in units of the hopping amplitude $J$, following a deep quench into the DW ordered phase for four repetitions of the quench simulation with random initial noise. The dynamics are modelled with variational Gaussian states and a lattice discretized model.}
\label{app:fig:gaussian}
\end{figure}

While linearized dynamics of the coupled atom-cavity system can be computed analytically and have good agreement with experiment, it's hard for it to predict how long the linear regime persists, and thus how many decades of exponential growth there are. To estimate the duration of the exponential regime, we employ variational Gaussian states \cite{SHI2018245} (GS). The variational parameters are the fermionic correlation matrix $\Gamma = \langle \psi_{\mathrm{GS}}\vert \hat{C}\hat{C}^{\dagger}\vert\psi_{\mathrm{GS}}\rangle$ and the photonic displacements $\Delta_\mathrm{R} = \langle \psi_{\mathrm{GS}}\vert (\hat{x},\hat{p})^T\vert\psi_{\mathrm{GS}}\rangle$ where $\vert\psi_{\mathrm{GS}}\rangle$ is the Gaussian wave function. In this way, we have the best possible description of the system up to two point correlators. We perform a 2D square lattice discretization of the Hamitonian (Eq.~\eqref{eq:hamiltonian}) with nearest neighbour hopping with hopping amplitude $J$ for the kinetic term. We work in two dimensions instead of three, since we do not expect solutions to change qualitatively for Gaussian states. We find the ground state for a given $U<0$ in the absence of light which is a BCS or BEC state, depending on the magnitude of $U$. We then apply a small noise transformation that preserves particle number to $\Gamma$ and $\Delta_R$ and perform real-time evolution with the light-matter coupling quenched above the critical value. The equations of motion are:

\begin{align}
    d_{t}\Delta_\mathrm{R} &= \sigma h_{\Delta},\label{general_realtime1}\\
    d_{t}\Gamma_\mathrm{f} &= i[\Gamma_\mathrm{f},h_\mathrm{f} ]. \nonumber\label{general_realtime3}
\end{align}

where $\sigma = \begin{pmatrix} 0 & 1\\-1&0\end{pmatrix}$ and:

\begin{align}
    h_{\Delta} &= 2\frac{\partial E}{\partial \Delta_\mathrm{R}},\\
    h_{f} &= -2\frac{\partial E}{\partial \Gamma_\mathrm{f}^\mathrm{T}} \nonumber \label{h_f_general}
\end{align}

which are the derivatives of the expectation value of energy $E = \langle \psi_{GS} \vert H\vert\psi_{GS}\rangle$ in the time-dependent Gaussian state. Energy and particle number are conserved in the simulations. For simplicity, we take a single ordering wave vector $\mathbf{q}=(4\pi/L_\mathrm{x}, 0)^T$. We expect translational invariance in the perpendicular direction. The results shown here are for a lattice of $L_\mathrm{x}\times L_\mathrm{y} = 16\times 4$ sites but the computation has favorable scaling for much larger studies in the future. We set the filling factor to $\nu=0.6$ so that the ratio $q/k_\mathrm{F}\approx0.4$ is close to the experimental value of $k_-$. We choose detuning $\tilde{\Delta}_{\mathrm{c}}=50J$ to have a good separation of scales. Fig.~\ref{app:fig:gaussian} shows intracavity photon number versus time for a deep quench from a BEC state to $D_0/D_{0\mathrm{C}} \gg 1$, with  $U=-2J$, for four simulations. The dissipation is set to $\kappa = 2J$. We observe several decades of uninterrupted exponential growth, which qualitatively agrees with experiment. We also check that reducing or increasing dissipation does not significantly alter the duration of the linear regime.

\newpage
\clearpage

\setcounter{figure}{0} 
\renewcommand{\thefigure}{E\arabic{figure}} 
\renewcommand{\figurename}{EXTENDED DATA FIG.}

\end{document}